\documentclass[12pt]{article}
\usepackage[usenames]{color}
\usepackage{setspace}
\usepackage{graphicx}
\usepackage{amsmath}
\usepackage{amssymb}
\usepackage{amsthm}
\usepackage{multirow}
\numberwithin{equation}{section}

\newcommand{\del}{\partial}

\newcommand{\bequ}{\begin{align}}
\newcommand{\eequ}{\end{align}}
\newcommand{\beqn}{\begin{align}}
\newcommand{\eeqn}{\end{align}}
\newcommand{\bctr}{\begin{center}}
\newcommand{\ectr}{\end{center}}
\newcommand{\bit}{\begin{itemize}}
\newcommand{\eit}{\end{itemize}}

\newcommand{\half}{{\frac12}}
\def\e{{\textrm e}}
\def\del{\partial}
\def\half{{\frac12}}

\def\del{\partial}
\def\half{{\frac12}}

\def\del{\partial}
\def\dslash{\del\kern-0.55em\raise 0.14ex\hbox{/}}

\def\rough#1{\raise.3ex\hbox{$#1$\kern-.75em\lower1ex\hbox{$\sim$}}}

\newcommand{\setTT}[6]{\substack{\{{#1},{#2},\cdots,{#3}\}\\ \in\{{#4},{#5},\cdots, {#6}\}}}

\newcommand{\PRD}[3]{{\it Phys. Rev.} {\bf D{#1}} (19{#3}) {#2}}

\newcommand{\PRDM}[3]{{\it Phys. Rev.} {\bf D{#1}} {#2} (20{#3})}

\newcommand{\PTP}[3]{{\it Prog. Theor. Phys.} {\bf {#1}} (19{#3}) {#2}}
\newcommand{\PTPM}[3]{{\it Prog. Theor. Phys.} {\bf {#1}} (20{#3}) {#2}}

\newcommand{\MPL}[3]{{\it Mod. Phys. Lett.} {\bf A{#1}} (19{#3}) {#2}}
\newcommand{\MPLM}[3]{{\it Mod. Phys. Lett.} {\bf A{#1}} (20{#3}) {#2}}

\newcommand{\jhep}[3]{{\it JHEP} {\bf {#1}} (20{#2}) {#3}}

\newcommand{\hepph}[1]{{\tt hep-ph/#1}}

\begin{document}
\begin{flushright}
{\small KOBE-TH-22-02}\\%
\end{flushright}
\begin{center}
{\LARGE\bf 
Non-analytic Term in Effective Potential \\
at Finite Temperature \\ 
for Scalar Field on Compactified Space \\
}
\vskip 1.4cm
Makoto Sakamoto$^{(a)}$
\footnote{E-mail: dragon@kobe-u.ac.jp} and
Kazunori Takenaga$^{(b)}$
\footnote{E-mail: takenaga@kumamoto-hsu.ac.jp}
\\
\vskip 1.0cm
${}^{(a)}$ {\it Department of Physics, Kobe University, 
Rokkodai Nada, Kobe, 657-8501 Japan}
\\[0.2cm]
${}^{(b)}$ {\it Faculty of Health Science, Kumamoto
Health Science University, Izumi-machi, Kita-ku, Kumamoto 861-5598, Japan}
\\
\vskip 1.5cm
\begin{abstract}
We study non-analytic terms, which cannot be written in the form of any positive integer power of 
field-dependent mass squared, in effective potential at finite temperature in one-loop approximation 
for a real scalar field on the $D$-dimensional space-time, $S_{\tau}^1\times R^{D-(p+1)}\times \prod_{i=1}^pS_i^1$.
The effective potential can be recast into the integral form in the complex plane by using the 
integral representation for the modified Bessel function of the second kind and the analytical
extension for multiple mode summations. The pole structure of the mode summations is 
clarified and all the non-analytic terms are obtained by the residue theorem. We find that the effective potential 
has a non-analytic term when the dimension of the flat Euclidean space, $D-(p+1)$ is odd. There appears only one 
non-analytic term  for the given values of $D$ and $p$, for which the non-analytic term exists.
\end{abstract}
\end{center}
\vskip 1.0 cm
\newpage
%
%
%
%
\section{Introduction}
%
%
%
%
The effective potential at finite temperature has provided a useful tool 
to study the phenomena of phase transition in quantum field theory. 
Dolan and Jackiw \cite{dj} found that there exists a non-analytic term, 
which cannot be written in the form of any positive integer power of field-dependent mass 
squared, in the effective potential at finite temperature for a scalar field. 
The non-analytic term found by them is proportional to three-halves power of 
the mass squared, and the term turns out to play a crucial role to trigger 
the first order phase transition\cite{quiros,dine}, for example, in electroweak theories. 
The magnitude of the term determines the strength of the first order phase transition 
and thus, it is concerned with the scenario of electroweak baryogenesis \cite{baryon} as well. 
Hence, the non-analytic term in the effective potential is an important quantity.

Quantum field theory with compactified dimensions has been one of the attractive approaches 
for physics beyond the standard model. 
Orbifold compactification, for example, provides an attractive framework for gauge-Higgs 
unification, where the Higgs field is unified into higher dimensional gauge fields 
and it is an alternative solution to the gauge hierarchy problem\cite{HIL,KLY}. 
The order of the phase transition in the gauge-Higgs unification at finite temperature 
has been studied in \cite{panico,marutake1}, and the first order phase transition 
can take place due to the term with three-halves power of the field-dependent mass squared 
in the effective potential. 
Compactified dimensions also offer the theoretical framework for studying 
quantum field theory itself. 
From a point of view of dimensional reduction \cite{cava,cava3}, models 
with several numbers of $S^1$ have been investigated. 
It has been also shown that the quantum field theory with compactified dimensions 
(at finite temperature) can possess rich phase structures \cite{HOST1,sakatake}.

Taking account of the aforementioned studies, it is important and interesting to investigate non-analytic terms in the
effective potential in the presence of extra dimensions at finite temperature. In this paper, we study all the non-analytic terms 
for a real scalar field on the $D$-dimensional space-time, $S_{\tau}^1\times R^{D-(p+1)}\times \prod_{i=1}^pS_i^1$, where $S_{\tau}^1$ stands 
for the Euclidean time direction and the spacial directions are compactified on $S_i^1$. The 
$R^{D-(p+1)}$ is the $D-(p+1)$ - dimensional flat Euclidean space. We assume that 
the scalar field satisfies the periodic boundary condition for the spacial $S^1$ direction.

The effective potential contains the modified Bessel function of the second kind accompanied with multiple mode summations. 
In addition to the integral representation for the modified Bessel function of the second kind given by the inverse Mellin
transformation \cite{bromwich,davis}, we make use of the analytical extension for the mode summations \cite{elizalde} in
order to recast the effective potential into the integral form in the complex plane and to
obtain the non-analytic terms by the residue theorem.

The analytical extension consists of the products of the gamma, zeta functions and their integrals.
We clarify the pole structure of the analytical extension and find that 
relevant terms in the mode summations for yielding the non-analytic terms satisfy a recurrence relation, which gives a general form for the relevant terms. 
Then, all the non-analytic terms can be
obtained by the residue theorem for the poles of the gamma and zeta functions in 
the general form, depending on the even/odd $D$ and $n~(n=1,2,\cdots, p+1)$. 
The positions of the poles that yield the non-analytic terms turn out not to depend on $D$.

We find that a non-analytic term appears in the effective potential when the dimension of the flat Euclidean 
space, $D-(p+1)$ is odd. There is only one non-analytic term  for the given values of $D$ and $p$, for which the non-analytic term exists.

This paper is organized as follows. In the next section, we rewrite the
effective potential in the integral form in the complex plane and discuss 
to obtain the non-analytic terms for the case of $p=1$ explicitly. 
And we obtain the general form for the relevant terms in the mode summations. 
We calculate the non-analytic terms by using the general form in the section $3$ 
and obtain the non-analytic terms in the effective potential in the section $4$. 
The final section is devoted to conclusions and discussions. 
Some details on the pole structure of the analytical extension in the mode summations are given in Appendix A.
%
%
%
%
%
%
%
%
\section{Effective potential in integral form and non-analytic terms}
%
%
We study non-analytic terms, which cannot be written in the form of any positive integer powers
of the field-dependent mass squared, in 
the effective potential for a real scalar field at finite temperature on the $D$-dimensional space-time,
$S_{\tau}^1\times R^{D-(p+1)}\times \prod_{i=1}^pS_i^1$ in one-loop approximation.
We employ the Euclidean time formalism for finite temperature field theory and then the
Euclidean time direction is compactified on $S_{\tau}^1$. The spacial $p$ dimensions are compactified on the $p$ numbers of 
$S^1$. We denote the circumference of each $S^1_{i}$ as $L_i (i=0, 1,\cdots, p)$ and $L_0$ stands for 
the inverse temperature $T^{-1}$.

One needs to evaluate 
\begin{align}
V_{\rm eff}
&=(-1)^{f}{\cal N}~\half\left( \prod_{i=0}^{p}\frac{1}{L_i}\sum_{n_i=-\infty}^{\infty}\right)
\int\frac{d^{D-(p+1)}p_E}{(2\pi)^{D-(p+1)}}
   \notag\\
&\hspace{5mm}\times
\log\Bigl[p_E^2 +\left(\frac{2\pi}{L_0}\right)^2(n_0 +\eta_0)^2
+\sum_{i=1}^p\left(\frac{2\pi}{L_i}\right)^2(n_i +\eta_i)^2
+M^2(\varphi)\Bigr]
\label{shiki2100}
\end{align}
in order to obtain the effective potential on 
$S_{\tau}^1\times R^{D-(p+1)}\times\prod_{i=1}^p S_i^1$ in one-loop approximation. 
The $M^2(\varphi)$ is the field-dependent mass squared of the scalar field \footnote{Hereafter, we denote $M(\varphi)$ by $M$ for simplicity.}. 
The $p_E$ denotes the $D-(p+1)$-dimensional Euclidean momentum. 
The $f$ is the fermion number, which is $0~(1)$ for bosons (fermions). 
The ${\cal N}$ is the on-shell degrees of freedom. 
The $n_{0}$ denotes the Matsubara mode at finite temperature and
the Kaluza-Klein mode $n_i~(i=1,\cdots, p)$ comes from each $S_i^1~(i=1,\cdots p)$.
The parameter $\eta_0$, which stands for the boundary condition for 
the $S_{\tau}^1$ direction, is determined by quantum statistics to be $0$ ($\half$) 
for bosons (fermions). The parameter $\eta_i~(i=1,\cdots, p)$ specifies the boundary condition
for the spacial $S_i^1$ direction.

We employ the zeta-function regularization in order to evaluate Eq.(\ref{shiki2100}). 
Let us define
%
\begin{align}
I(s)&\equiv \left( \prod_{i=0}^{p}\frac{1}{L_i}\sum_{n_i=-\infty}^{\infty}\right)
\int\frac{d^{D-(p+1)}p_E}{(2\pi)^{D-(p+1)}}\notag\\
&\hspace{5mm}
\times 
\Bigl[p_E^2 +\left(\frac{2\pi}{L_0}\right)^2(n_0 +\eta_0)^2
+\sum_{i=1}^p\left(\frac{2\pi}{L_i}\right)^2(n_i +\eta_i)^2
+M^2\Bigr]^{-s}.
\end{align}
%
Then, the effective potential can be written as
%
\begin{equation}
V_{\rm eff}=(-1)^{f}\frac{\cal N}{2}\left(-\frac{d}{ds}I(s)\right)\Bigg|_{s\rightarrow 0}.
\end{equation}
%
Using the formula 
%
\begin{align}
A^{-s} = \frac{1}{\Gamma(s)} \int_{0}^{\infty}\!dt\,t^{s-1} e^{-At}
\label{new2.4}
\end{align}
%
and the Poisson summation
(\ref{eqA.2}) in Appendix A with the replacement of 
$L_{0} \to \frac{2\pi}{L_{i}}$, $m_{0} \to n_{i}$, $n_{0} \to m_{i}$ and
$\eta_{0} \to \eta_{i}$,
we arrive at
%
\begin{align}
V_{\rm eff}&=(-1)^{f+1}\frac{\cal N}{2}\frac{\pi^{\frac{D}{2}}}{(2\pi)^D}
\sum_{m_0=-\infty}^{\infty}\cdots \sum_{m_p=-\infty}^{\infty}
\notag \\
&\hspace{5mm}\times
  \int_0^{\infty}dt~t^{-\frac{D}{2}-1}\e^{-\frac{1}{4t}[(m_0L_0)^2+\cdots+(m_pL_p)^2]
  -M^2t +2\pi i(m_0\eta_0+\cdots +m_p\eta_p)}.
\label{eq2.4}
\end{align}
%
%

%
It is convenient to separate each summation $\sum_{m_{i}}$ in Eq.(\ref{eq2.4}) 
into the zero mode ($m_{i}=0$) and the non-zero ones ($m_{i} \ne 0$), and
to express Eq.(\ref{eq2.4}) into the form
%
\begin{align}
V_{\text{eff}} = \sum_{n=0}^{p+1} F^{(n)},
\label{shikinew1_1}
\end{align}
%
where
%
\begin{align}
F^{(n)} 
 &= \sum_{0\leq i_1<i_2<\cdots < i_n\leq p}F^{(n)}_{L_{i_1},L_{i_2},\cdots, L_{i_n}},
    \label{shikinewTneq0}\\
F^{(n)}_{L_{i_1},L_{i_2},\cdots, L_{i_n}}
 &= (-1)^{f+1}\frac{\cal N}{2}\frac{\pi^{\frac{D}{2}}}{(2\pi)^D}
     \sum_{m_{i_1}=-\infty}^{\infty}{}^{\hspace{-3mm}\prime}\cdots 
     \sum_{m_{i_n}=-\infty}^{\infty}{}^{\hspace{-3mm}\prime}
     \notag \\
 &\hspace{5mm}\times
    \int_0^{\infty}dt~t^{-\frac{D}{2}-1}\e^{-\frac{1}{4t}[(m_{i_1}L_{i_1})^2+
     \cdots+(m_{i_n}L_{i_n})^2]-M^2t +2\pi i(m_{i_1}\eta_{i_1}+\cdots +m_{i_n}\eta_{i_n})}.
\end{align}
%
Here, the prime of the summation $\sum_{m_{i}=-\infty}^{\,\prime\, \infty}$
means that the zero mode ($m_{i}=0$) is removed.

The $F^{(0)}$ in Eq.(\ref{shikinew1_1}) corresponds to the contribution from
all the zero modes $m_{0} = m_{1} = \cdots = m_{p} = 0$ in Eq.(\ref{eq2.4})
and is given by \footnote{It must be understood that $F^{(0)}$ is regularized by 
the dimensional regularization for $D=$ even.}
%
\begin{align}
F^{(0)}
 = (-1)^{f+1}\frac{\cal N}{2}\frac{\pi^{\frac{D}{2}}}{(2\pi)^D}
    \int_0^{\infty}dt~t^{-  \frac{D}{2}-1}\e^{-M^2t}
 = (-1)^{f+1}\frac{\cal N}{2}\frac{\pi^{\frac{D}{2}}}{(2\pi)^D}
   \Gamma(-\tfrac{D}{2})(M^2)^{\frac{D}{2}}.
\label{shiki2102}
\end{align}
%
On the other hand, by using the formula (\ref{eqA.3}) in Appendix A, 
$F^{(n)}_{L_{i_1},L_{i_2},\cdots, L_{i_n}}\ (n \ge 1)$ can be obtained as
%
\begin{align}
F_{L_{i_1},L_{i_2},\cdots, L_{i_{n}}}^{(n)}
 &= (-1)^{f+1}{\cal N}\frac{2^n}{(2\pi)^{\frac{D}{2}}}
    \sum_{m_{i_{1}}=1}^{\infty} \cdots \sum_{m_{i_{n}}=1}^{\infty}
    \left(\frac{M^2}{(m_{i_1}L_{i_1})^2+\cdots +(m_{i_n}L_{i_n})^2}\right)^{\frac{D}{4}}
    \notag\\
&\hspace{5mm}\times 
    K_{\frac{D}{2}}\left(\sqrt{{M^2}\{(m_{i_1}L_{i_1})^2+\cdots +(m_{i_n}L_{i_n})^2\}}\right)
    \cos(2\pi m_{i_1}\eta_{i_1})\cdots \cos(2\pi m_{i_n}\eta_{i_n}).
\label{shikinewgen}
\end{align}
%
In this paper, we consider a real  scalar field ($f=0, {\cal N}=1$) and take 
the periodic boundary condition $\eta_{j}=0~(j=1,\cdots, p)$  
for the spacial $S^1_{j}$ direction.

\subsection{Non-analytic terms for the case of $S_{\tau}^1\times R^{D-2}\times S_1^1$}
%
%
%
Let us first consider that the space-time is $S_{\tau}^1\times R^{D-2}\times S_1^1$ and study non-analytic terms in the
effective potential. 
Although the results concerning the effective potential in this 
subsection are not new and, in fact, they are given in \cite{cava}, 
this is a simple but nontrivial example 
and is appropriate to present here the analysis on the non-analytic terms in detail.

The effective potential on $S^{1}_{\tau}\times R^{D-2} \times S^{1}_{1}$
takes the form
\begin{align}
V_{\rm eff}=F^{(0)}+F^{(1)} +F^{(2)},
\label{shiki2101}
\end{align}
where $F^{(0)}$ is given by Eq.(\ref{shiki2102}) and 
\begin{align}
F^{(1)}&= F_{L_0}^{(1)}+F_{L_1}^{(1)}\nonumber\\
&=
-\frac{2}{(2\pi)^{\frac{D}{2}}}
\sum_{m_0=1}^{\infty}\left(\frac{M^2}{m_0^2L_0^2}\right)^{\frac{D}{4}}
K_{\frac{D}{2}}(\sqrt{M^2(m_0L_0)^2})
\nonumber\\
&\hspace{5mm}
-\frac{2}{(2\pi)^{\frac{D}{2}}}
\sum_{m_1=1}^{\infty}\left(\frac{M^2}{m_1^2L_1^2}\right)^{\frac{D}{4}}
K_{\frac{D}{2}}(\sqrt{M^2(m_1L_1)^2}), 
\label{shiki2103}\\
F^{(2)}&= 
F_{L_0,L_1}^{(2)}
\nonumber\\
&=
-\frac{2^2}{(2\pi)^{\frac{D}{2}}}\sum_{m_0=1}^{\infty}\sum_{m_1=1}^{\infty}
\left(\frac{M^2}{(m_0L_0)^2+(m_1L_1)^2}\right)^{\frac{D}{4}}
K_{\frac{D}{2}}\left(\sqrt{M^2\{(m_0L_0)^2+(m_1L_1)^2\}}\right),\nonumber\\
\label{shiki2104}
\end{align}
which follow from Eq.(\ref{shikinewgen}).
Here, the $K_{\frac{D}{2}}(x)$ is the modified Bessel function of the second kind. 
Let us note that $F^{(0)} + F_{L_0}^{(1)}$ is the effective
potential at finite temperature without the compactified spacial dimension.

For our purpose, let us use the integral representation for 
the modified Bessel function of the second kind \cite{bromwich} in the complex plane 
\footnote{This is the inverse Mellin transformation for the $K_{\nu}(x)$\cite{davis}.},
\begin{align}
K_{\nu}(x)=\frac{1}{4\pi i}\int_{c-i\infty}^{c+i\infty}dt~\Gamma(t)\Gamma(t-\nu)\left(\frac{x}{2}\right)^{-2t +\nu} .
\label{shiki2105}
\end{align}
The constant $c$ should be understood to be a point located on the real axis which is greater than all the poles of 
the gamma functions in the integrand. Then, we deform the integration path in such a way that it encloses all the poles 
in the integrand and we can perform the $t$-integration by the residue theorem.

If we apply Eq.(\ref{shiki2105}) to the firsr term of Eq.(\ref{shiki2103}) , we have
\begin{align}
F^{(1)}_{L_0}=
-\frac{2}{(2\pi)^{\frac{D}{2}}}\left(\frac{M^2}{2}\right)^{\frac{D}{2}}\frac{1}{4\pi i}
\int_{c-i\infty}^{c+i\infty}dt~\Gamma(t)\Gamma(t-\tfrac{D}{2})\zeta(2t)
\left(\frac{ML_0}{2}\right)^{-2t}.
\label{shiki2106}
\end{align}
The zeta function $\zeta(2t)$ is the consequence of the single mode summation with respect to $m_0$. One 
can obtain all the terms in the effective potential in terms of $M$ 
\cite{quiros,braden}
by the residue integral for all the poles in the integrand.

Once we obtain the integral form like Eq.(\ref{shiki2106}), it is easy to find the 
non-analytic terms, which cannot be written in the form of any positive integer powers of 
$M^2$. 
The pole at $t=\frac{D}{2}-n~(n=0,1,2,\cdots)$ of $\Gamma(t-\frac{D}{2})$ yields
the mass dependence on $(M^2)^{\frac{D}{2}}M^{-2t}=(M^2)^n$, 
so that the residue integral for the pole 
does not produce non-analytic terms. It turns out that any poles depending on $D$ do not produce non-analytic terms.
This observation is crucially used throughout our discussions.

Moreover, we find that the mass dependence on $(M^{2})^{\frac{D}{2}} M^{-2t}$
in Eq.(\ref{shiki2106}) tells us that for $D=$ even (odd), half-odd-integer
(integer) values of $t$ can yield non-analytic terms.
It follows that a non-analytic term in Eq.(\ref{shiki2106}) arises from
either a pole of $\Gamma(t)$ at $t=0$ for $D=$ odd or that of $\zeta(2t)$
at $t=\frac{1}{2}$ for $D=$ even, whose value of the pole is independent of $D$, as stated above.
One might think that the poles of $\Gamma(t)$ at $t = -n\ (n=1,2,\cdots)$
could produce non-analytic terms for $D=$ odd.
This is not, however, the case because of the property $\zeta(2t) = 0$
for $t = -n\ (n=1,2,\cdots)$.
This observation is also used throughout our discussions.

For $D=$ even, the residue theorem for the pole $t=\half$ of $\zeta(2t)$ gives us the 
non-analytic term given by
\begin{align}
F^{(1)}_{L_0}\Big|_{\rm n.a.}=
-\frac{(-1)^{\frac{D}{2}}}{2^{\frac{D}{2}}\pi^{\frac{D-2}{2}}(D-1)!!}\frac{M^{D-1}}{L_0},
\label{shiki2107}
\end{align}
where we have used 
\begin{align}
\Gamma(\tfrac{1-D}{2})\Big|_{D={\rm even}}
=\frac{(-1)^{\frac{D}{2}}2^{\frac{D}{2}}}{(D-1)!!}\sqrt{\pi}.
\label{shiki2108}
\end{align}
The abbreviation denoted by ``{\rm n.a.}'' in Eq.(\ref{shiki2107}) means {\it non-analytic terms}.
Eq.(\ref{shiki2107}) is the famous term found by Dolan and Jackiw \cite{dj} for $D=4$. 
The other poles of the gamma functions in the integrand and Eq.(\ref{shiki2102}) for $D=$ even do not 
yield non-analytic terms, so that the Dolan-Jackiw term (\ref{shiki2107}) is the only 
possible non-analytic term in the effective potential for $D=4$ and $p=0$.

For $D=$ odd, on the other hand, the pole $t=0$ of $\Gamma(t)$ in Eq.(\ref{shiki2106}) gives us 
\begin{align}
F^{(1)}_{L_0}\Big|_{\rm n.a.}=\frac{(-1)^{\frac{D+1}{2}}}{2^{\frac{D+1}{2}}\pi^{\frac{D-1}{2}}D!!}M^D ,
\label{shiki2109}
\end{align}
where we have used 
\begin{align}
\Gamma(-\tfrac{D}{2})\Big|_{D={\rm odd}}
=\frac{(-1)^{\frac{D+1}{2}}2^{\frac{D+1}{2}}}{D!!}\sqrt{\pi}  .
\label{shiki2110}
\end{align}
Eq.(\ref{shiki2109}), however, cancels the non-analytic term of Eq.(\ref{shiki2102}). 
Thus, there is no non-analytic term in the effective potential for $D=$ odd and $p=0$.

Likewise, one can evaluate the second term of Eq.(\ref{shiki2103}) and obtains the non-analytic term as
\begin{align}
F^{(1)}_{L_1}\Big|_{\rm n.a.}=
\left\{\begin{array}{ll}
\displaystyle{
-\frac{(-1)^{\frac{D}{2}}}{2^{\frac{D}{2}}\pi^{\frac{D-2}{2}}(D-1)!!}\frac{M^{D-1}}{L_1}}&\quad {\rm for}\quad D={\rm even},
\\[0.5cm]
\displaystyle{
\frac{(-1)^{\frac{D+1}{2}}}{2^{\frac{D+1}{2}}\pi^{\frac{D-1}{2}}D!!}M^D}&\quad {\rm for}\quad D={\rm odd} .
\end{array}\right.
\label{shiki2113}
\end{align}

Let us next proceed to study non-analytic terms in $F^{(2)}=F_{L_0,L_1}^{(2)}$, which is rewritten in the integral form 
by using Eq.(\ref{shiki2105}) as
\begin{align}
F_{L_0,L_1}^{(2)}
&=-\frac{2^2}{(2\pi)^{\frac{D}{2}}}\left(\frac{M^2}{2}\right)^{\frac{D}{2}}
\frac{1}{4\pi i}\int_{c-i\infty}^{c+i\infty}dt~\Gamma(t-\tfrac{D}{2})\left(\frac{M}{2}\right)^{-2t}\nonumber\\
&\hspace{5mm}
  \times \Gamma(t)\sum_{m_0=1}^{\infty}\sum_{m_1=1}^{\infty}\bigl\{(m_0L_0)^2+(m_1 L_1)^2\bigr\}^{-t}  .
\label{shiki2114}
\end{align}
One must evaluate the double summations
\begin{align}
\Gamma(t)
\sum_{m_0=1}^{\infty}\sum_{m_1=1}^{\infty}\bigl\{(m_0L_0)^2+(m_1 L_1)^2\bigr\}^{-t},
\label{shiki2115}
\end{align}
which is known to have an analytical extension \cite{elizalde}, as shortly discussed in
Appendix A, given by (see Eq.(\ref{eqA.4}))
\begin{align}
&\Gamma(t) \sum_{m_0=1}^{\infty}\sum_{m_1=1}^{\infty}\bigl\{(m_0L_0)^2+(m_1 L_1)^2\bigr\}^{-t} \nonumber\\
&=-\half \frac{1}{L_1^{2t}}\Gamma(t)\zeta(2t) + \frac{\sqrt{\pi}}{2}
\frac{1}{L_0L_1^{\,2t-1}}\Gamma(t-\tfrac{1}{2})\zeta(2t -1)\nonumber\\
&\hspace{5mm}
+\frac{1}{\sqrt{\pi}}\left(\frac{\pi}{L_0}\right)^{2t}\frac{1}{2\pi i}
\int_{c_{1} -i\infty}^{c_{1}+i\infty}d{t_{1}}~\Gamma(t_{1} -t +\tfrac{1}{2})
\zeta(2t_{1} -2t +1)\Gamma(t_{1})\zeta(2t_{1})\left(\pi\frac{L_1}{L_0}\right)^{-2t_{1}},
\label{shiki2117_1}\nonumber\\
\end{align}
where we have first summed over $m_0$ and then $m_1$.

Inserting (\ref{shiki2117_1}) into (\ref{shiki2114}), we have
\begin{align}
F_{L_0,L_1}^{(2)}
&=\frac{1}{(2\pi)^{\frac{D}{2}}}\left(\frac{M^2}{2}\right)^{\frac{D}{2}}\frac{1}{2\pi i}
\int_{c -i\infty}^{c+i\infty}dt~\Gamma(t-\tfrac{D}{2})\Gamma(t)
\zeta(2t)\left(\frac{ML_1}{2}\right)^{-2t}\nonumber\\
&\hspace{5mm}
-\frac{1}{(2\pi)^{\frac{D}{2}}}{\sqrt{\pi}}\left(\frac{M^2}{2}\right)^{\frac{D}{2}}\frac{L_1}{L_0}\frac{1}{2\pi i}
\int_{c -i\infty}^{c+i\infty}dt~\Gamma(t-\tfrac{D}{2})
\Gamma(t-\tfrac{1}{2})\zeta(2t-1)\left(\frac{ML_1}{2}\right)^{-2t}\nonumber\\
&\hspace{5mm}
-\frac{2}{(2\pi)^{\frac{D}{2}}}\left(\frac{M^2}{2}\right)^{\frac{D}{2}}
\frac{1}{2\pi i}\int_{c -i\infty}^{c+i\infty}dt~\Gamma(t-\tfrac{D}{2})
\left(\frac{M}{2}\right)^{-2t}\frac{1}{\sqrt{\pi}}
\left(\frac{\pi}{L_{0}}\right)^{2t}
\nonumber\\
&
\hspace{10mm}
\times
\frac{1}{2\pi i}
\int_{c_{1} -i\infty}^{c_{1}+i\infty}d{t_{1}}~\Gamma(t_{1} -t +\tfrac{1}{2})
\zeta(2t_{1} -2t +1)\Gamma(t_{1})\zeta(2t_{1})\left(\pi\frac{L_1}{L_0}\right)^{-2t_{1}}.
\label{shiki2118}
\end{align}
We note again that the poles of $\Gamma(t-\frac{D}{2})$ do not yield non-analytic terms.
This is because the residue integral with the poles of $\Gamma(t-\frac{D}{2})$
at $t = \frac{D}{2} - n\ (n=0,1,2,\cdots)$ gives the mass dependence on 
$(M^{2})^{\frac{D}{2}}M^{-2t} = (M^{2})^{n}$,
which are not non-analytic terms for $n=0,1,2,\cdots$.
Thus, the poles, whose positions depend on $D$, do not give non-analytic terms,
as mentioned before.

We find that non-analytic terms in Eq.(\ref{shiki2118}) can arise only from
the poles other than those of $\Gamma(t-\frac{D}{2})$.
In fact, the non-analytic terms are produced by the poles of $t=0$ for $D=$ odd
and $t=\frac{1}{2}$ for $D=$ even ($t=\frac{1}{2}$ for $D=$ even and $t=1$ for
$D=$ odd) in the first (second) term of Eq.(\ref{shiki2118}).

On the other hand, there is no pole that contributes to the residue integral with respect to $t$
in the third term of Eq.(\ref{shiki2118}), other than those of $\Gamma(t-\frac{D}{2})$,
as shown in the subsection A.1 of Appendix A.
This implies that the third term in Eq.(\ref{shiki2118}) is irrelevant for the
analyses of the non-analytic terms because the poles of $\Gamma(t-\frac{D}{2})$ do not
produce any non-analytic terms, as stressed above.
This is a crucial observation for our study on the non-analytic terms in the
effective potential and plays a central role in the discussions.

The non-analytic terms in $F_{L_0,L_1}^{(2)}$ turn out to come from the first and the second terms in Eq.(\ref{shiki2118}) 
and we obtain 
\begin{align}
F_{L_0, L_1}^{(2)}\Big|_{\rm n.a.}=
\frac{(-1)^{\frac{D}{2}}}{2^{\frac{D}{2}}\pi^{\frac{D-2}{2}}(D-1)!!}M^{D-1}
\left(\frac{1}{L_1}+\frac{1}{L_0}\right)
\label{shiki2119}
\end{align}
for $D=$ even. 
The first (second) term in Eq.(\ref{shiki2119}) comes from the 
pole $t=\half$ of $\zeta(2t)$ ($t=\half$ of $\Gamma(t-\half)$) in the first (second) term of 
Eq.(\ref{shiki2118}).
%
%
On the other hand, for $D=$ odd, we obtain the non-analytic terms as
\begin{align}
F_{L_0,L_1}^{(2)}\Big|_{\rm n.a.}=
-\frac{(-1)^{\frac{D+1}{2}}}{2^{\frac{D+1}{2}}\pi^{\frac{D-1}{2}}D!!}M^D
-\frac{(-1)^{\frac{D-1}{2}}}{2^{\frac{D-1}{2}}\pi^{\frac{D-3}{2}}(D-2)!!}\frac{M^{D-2}}{L_0L_1}.
\label{shiki2120}
\end{align}
The first (second) term of Eq.(\ref{shiki2120}) comes from the pole  
$t=0$ of $\Gamma(t)$ ($t=1$ of $\zeta(2t-1)$) in the first (second) term of Eq.(\ref{shiki2118}).

Collecting the terms we have obtained above, we find that for $D=$ even 
the effective potential has no non-analytic term, i.e.
\begin{align}
V_{\rm eff}\Big|_{\rm n.a.}=(F^{(0)}+F^{(1)}_{L_0}+F^{(1)}_{L_1} +F_{L_0,L_1}^{(2)})\Big|_{\rm n.a.} = 0
\qquad \text{for $D=$ even}\,,
\label{shiki2121}
\end{align}
although $F^{(1)}_{L_{0}}, F^{(1)}_{L_{1}}$ and $F^{(2)}_{L_{0},L_{1}}$
contain the non-analytic terms.
For $D=$ odd, all of 
$F^{(0)}, F^{(1)}_{L_{0}}, F^{(1)}_{L_{1}}$ and $F^{(2)}_{L_{0},L_{1}}$
have the non-analytic terms, but some of them cancel each other.
Then, the result is given by
\begin{align}
V_{\rm eff}\Big|_{\rm n.a.}
&=(F^{(0)}+F^{(1)}_{L_0}+F^{(1)}_{L_1} +F_{L_0,L_1}^{(2)})\Big|_{\rm n.a.} 
    \notag\\
&= 
-\frac{(-1)^{\frac{D-1}{2}}}{2^{\frac{D-1}{2}}\pi^{\frac{D-3}{2}}(D-2)!!}\frac{M^{D-2}}{L_0L_1} \qquad \text{for $D=$ odd}.
\label{shiki2122}
\end{align}
We conclude that the effective potential for the scalar field on 
$S_{\tau}^1\times R^{D-2}\times S_1^1$ 
does not have non-analytic terms for $D=$ even, while for $D=$ odd it has the non-analytic term 
given by Eq.(\ref{shiki2122}).
%
%
%
\subsection{General form for relevant terms in mode summation}
%
%
%
In this subsection, we investigate 
$F^{(n)}_{L_{0},L_{1},\cdots,L_{n-1}}\ (n=1,2,\cdots,p+1)$
and rewrite it in a tractable form to obtain the non-analytic terms.
By use of the formula (\ref{shiki2105}), Eq.(\ref{shikinewgen})
with $f=0, {\cal N}=1$ and $\eta_{i}=0$ can be written as
%
\begin{align}
&F^{(n)}_{L_{0},L_{1},\cdots,L_{n-1}}
  \notag\\
&= -\frac{2^{n}}{(2\pi)^{\frac{D}{2}}}\bigg(\frac{M^{2}}{2}\bigg)^{\frac{D}{2}}
   \frac{1}{4\pi i} \int_{c-i\infty}^{c+i\infty}dt\,\Gamma(t-\tfrac{D}{2})
   \bigg(\frac{M}{2}\bigg)^{-2t}
   \notag\\
&\hspace{5mm}
  \times \Gamma(t) \sum_{m_{0}=1}^{\infty}\sum_{m_{1}=1}^{\infty}\cdots
  \sum_{m_{n-1}=1}^{\infty}
  \Big\{ (m_{0}L_{0})^{2}+(m_{1}L_{1})^{2}+\cdots(m_{n-1}L_{n-1})^{2} \Big\}^{-t}
  \notag\\
&= -\frac{2^{n}}{(2\pi)^{\frac{D}{2}}}\bigg(\frac{M^{2}}{2}\bigg)^{\frac{D}{2}}
   \frac{1}{4\pi i} \int_{c-i\infty}^{c+i\infty}dt\,\Gamma(t-\tfrac{D}{2})
   \bigg(\frac{M}{2}\bigg)^{-2t} S^{(n)}(t; L_{0}, L_{1}, \cdots, L_{n-1})\,,
\label{eq2.27}
\end{align}
%
where we have defined
%
\begin{align}
&S^{(n)}(t; L_{0}, L_{1}, \cdots, L_{n-1})
  \notag\\
&\hspace{5mm}
 \equiv \Gamma(t) \sum_{m_{0}=1}^{\infty}\sum_{m_{1}=1}^{\infty}\cdots
  \sum_{m_{n-1}=1}^{\infty}
   \Big\{ (m_{0}L_{0})^{2}+(m_{1}L_{1})^{2}+\cdots(m_{n-1}L_{n-1})^{2} \Big\}^{-t}.
\label{eq2.28}
\end{align}
%
%

Inserting the formula (\ref{eqA.8}) into Eq.(\ref{eq2.27}), we have
%
\begin{align}
&F^{(n)}_{L_{0},L_{1},\cdots,L_{n-1}}
  \notag\\
&= -\frac{2^{n}}{(2\pi)^{\frac{D}{2}}}\bigg(\frac{M^{2}}{2}\bigg)^{\frac{D}{2}}
   \frac{1}{4\pi i} \int_{c-i\infty}^{c+i\infty}dt\,\Gamma(t-\tfrac{D}{2})
   \bigg(\frac{M}{2}\bigg)^{-2t}
   \notag\\
&\hspace{5mm}
  \times 
   \bigg\{
     -\frac{1}{2} S^{(n-1)}(t;L_{1},L_{2},\cdots,L_{n-1})
      + \frac{\sqrt{\pi}}{2L_{0}}S^{(n-1)}(t-\tfrac{1}{2};L_{1},L_{2},\cdots,L_{n-1})
       \notag\\
&\hspace{10mm}
  + \frac{1}{\sqrt{\pi}} 
    \frac{1}{2\pi i} \int_{c_1-i\infty}^{c_1+i\infty}dt_1\,
    \Gamma(t_{1}-t+\tfrac{1}{2}) \zeta(2t_{1}-2t+1)
    \bigg(\frac{\pi}{L_{0}}\bigg)^{-2t_{1}+2t}
    S^{(n-1)}(t_{1};L_{1},L_{2},\cdots,L_{n-1})
   \bigg\} \,.
\label{eq2.29}
\end{align}
%
Since we are interested in the non-analytic terms of the effective potential,
Eq.(\ref{eq2.29}) can be expressed as
%
\begin{align}
&F^{(n)}_{L_{0},L_{1},\cdots,L_{n-1}}\Big|_{\textrm{n.a.}}
  \notag\\
&= -\frac{2^{n}}{(2\pi)^{\frac{D}{2}}}\bigg(\frac{M^{2}}{2}\bigg)^{\frac{D}{2}}
   \frac{1}{4\pi i} \int_{c-i\infty}^{c+i\infty}dt\,\Gamma(t-\tfrac{D}{2})
   \bigg(\frac{M}{2}\bigg)^{-2t}
   \notag\\
&\hspace{5mm}
  \times 
   \bigg\{
     -\frac{1}{2} S^{(n-1)}(t;L_{1},L_{2},\cdots,L_{n-1})
      + \frac{\sqrt{\pi}}{2L_{0}}S^{(n-1)}(t-\tfrac{1}{2};L_{1},L_{2},\cdots,L_{n-1})
   \bigg\}\bigg|_{\textrm{n.a.}} \,,
\label{eq2.30}
\end{align}
%
where we have dropped the third term in Eq.(\ref{eq2.29}) because it is irrelevant
to obtain the non-analytic terms in $F^{(n)}_{L_{0},L_{1},\cdots,L_{n-1}}$
due to the fact that only the poles of $\Gamma(t-\frac{D}{2})$ contribute to
the residue integral with respect to $t$ in the third term of Eq.(\ref{eq2.29}), 
as shown in the subsection A.2 of Appendix A,
and those of $\Gamma(t-\frac{D}{2})$ do not produce any non-analytic terms,
as stressed before. This is a crucial observation in our analysis.

It follows from Eq.(\ref{eq2.27}) and Eq.(\ref{eq2.30}) that as long as our considerations
are restricted to the non-analytic terms, we can use the recurrence relation
%
\begin{align}
&S^{(n)}(t;L_{0},L_{1},\cdots,L_{n-1})
  \notag\\
&\hspace{5mm}  
 = -\frac{1}{2} S^{(n-1)}(t;L_{1},L_{2},\cdots,L_{n-1})
      + \frac{\sqrt{\pi}}{2L_{0}}S^{(n-1)}(t-\tfrac{1}{2};L_{1},L_{2},\cdots,L_{n-1})
\label{eq2.31}
\end{align}
%
in Eq.(\ref{eq2.30}).
The recurrence relation (\ref{eq2.31}) is easily solved as
%
\begin{align}
&S^{(n)}(t;L_{0},L_{1},\cdots,L_{n-1})
  \notag\\
&\hspace{5mm} 
  = \frac{(-1)^{n-1}}{2^{n-1}} \sum_{\ell=0}^{n-1}(-1)^{\ell}\pi^{\tfrac{\ell}{2}}
   \sum_{0 \le i_{1} < i_{2} < \cdots < i_{\ell} \le n-2}
    \big(L_{i_{1}}L_{i_{2}} \cdots L_{i_{\ell}}\big)^{-1}
     S^{(1)}(t-\tfrac{\ell}{2};L_{n-1})\,,
\label{eq2.32}
\end{align}
%
in terms of $S^{(1)}(t-\frac{\ell}{2};L_{n-1})$.
By direct calculations, $S^{(1)}(t;L)$ is found to be 
%
\begin{align}
S^{(1)}(t;L) = \Gamma(t) \sum_{m=1}^{\infty}\big\{(mL)^{2}\big\}^{-t}
             = \frac{\Gamma(t)\zeta(2t)}{L^{2t}}\,.
\label{eq2.33}
\end{align}
%
Thus, we have found that
%
\begin{align}
&F^{(n)}_{L_{0},L_{1},\cdots,L_{n-1}} \Big|_{\textrm{n.a.}}
  \notag\\
&= \frac{(-1)^{n}}{(2\pi)^{\frac{D}{2}}}\bigg(\frac{M^{2}}{2}\bigg)^{\frac{D}{2}}
   \frac{1}{2\pi i} \int_{c-i\infty}^{c+i\infty}dt\,\Gamma(t-\tfrac{D}{2})
   \bigg(\frac{ML_{n-1}}{2}\bigg)^{-2t}
   \notag\\
&\hspace{5mm}
  \times 
   \sum_{\ell=0}^{n-1}(-1)^{\ell}\pi^{\frac{\ell}{2}}
   \sum_{0 \le i_{1} < i_{2} < \cdots < i_{\ell} \le n-2}
    \big(L_{i_{1}}L_{i_{2}} \cdots L_{i_{\ell}}\big)^{-1} (L_{n-1})^{\ell}
    \Gamma(t-\tfrac{\ell}{2}) \zeta(2t-\ell) \bigg|_{\textrm{n.a.}}\,,
\label{eq2.34}
\end{align}
%
which will be used in the next section.
%
%
%
%
%
\section{Non-analytic terms in $F_{L_0,L_1,\cdots, L_{n-1}}^{(n)}$}
%
%
%
We are ready to calculate the non-analytic terms by using
Eq.(\ref{eq2.34})
for each case of even/odd $D$ and $n$. 
We note again that
the mass dependence of $(M^2)^{\frac{D}{2}-t}$ in Eq.(\ref{eq2.34}) 
tells us that $t=$ half-odd-integer (integer) poles 
of $\Gamma(t-\frac{\ell}{2})$ and $\zeta(2t-\ell)$ yield the non-analytic terms 
for $D=$ even (odd). 
%
%
%
%
\subsection{$(D, n)=({\rm even}, {\rm even})$}
%
%
%
Since $D=$ even in the present case, one needs the poles which are half-odd-integers 
in the integrand of 
Eq.(\ref{eq2.34}) in order to
have the non-analytic terms, that is, the poles of $\Gamma(t-\frac{\ell}{2})$ for $\ell=$ odd and those of the 
$\zeta(2t-\ell)$ for $\ell=$ even. 
Hence, it is convenient to separate the summation over $\ell$ into $\ell=$ odd and $\ell=$ even, as follows:
\begin{align}
&F_{L_0,L_1,\cdots, L_{n-1}}^{(n)}\Big|_{\rm n.a.}
=\frac{1}{(2\pi)^{\frac{D}{2}}}
\left(\frac{M^2}{2}\right)^{\frac{D}{2}}
\frac{1}{2\pi i}\int_{c-i\infty}^{c+i\infty}dt~\Gamma(t-\tfrac{D}{2})\left(\frac{ML_{n-1}}{2}\right)^{-2t}
\nonumber\\
&\times
\Biggl( 
\sum_{j=1}^{\frac{n}{2}}(-1)^{2j-1}\pi^{\frac{2j-1}{2}}
\hspace{-8mm}
\sum_{0\leq i_1<i_2<\cdots <i_{2j-1}\leq n-2}
\hspace{-8mm}
(L_{i_1}L_{i_2}\cdots L_{i_{2j-1}})^{-1}(L_{n-1})^{2j-1}
\Gamma(t-\tfrac{2j-1}{2})\zeta(2t-(2j-1))\nonumber\\
&\hspace{8mm}
+
\sum_{j=0}^{\frac{n}{2}-1}(-1)^{2j}\pi^{j}
\hspace{-8mm}\sum_{0\leq i_1<i_2<\cdots <i_{2j}\leq n-2}
\hspace{-8mm}
(L_{i_1}L_{i_2}\cdots L_{i_{2j}})^{-1}(L_{n-1})^{2j}
\Gamma(t-j)\zeta(2t-2j)\Biggr)\Bigg|_{\rm n.a.}.
\label{newshiki11}
\end{align}
The residue integral for the first term in Eq.(\ref{newshiki11}) is performed by the pole $t=\frac{2j-1}{2}$ of the gamma 
function, which yields
\begin{align}
\sum_{j=1}^{\frac{n}{2}}
\frac{(-1)^{\frac{D}{2}}(-1)^{j-1}M^{D-(2j-1)}}{2^{\frac{D-2(j-1)}{2}}\pi^{\frac{D-2j}{2}}(D-(2j-1))!!}
\sum_{0\leq i_1<i_2<\cdots <i_{2j-1}\leq n-2}
(L_{i_1}L_{i_2}\cdots L_{i_{2j-1}})^{-1}.
\label{newshiki12}
\end{align}
The pole $t=\frac{2j+1}{2}$ of the zeta function in the second term of Eq.(\ref{newshiki11}) gives us
\begin{align}
&\sum_{j=0}^{\frac{n}{2}-1}
\frac{(-1)^{\frac{D}{2}}(-1)^{j}M^{D-(2j+1)}}{2^{\frac{D-2j}{2}}\pi^{\frac{D-2(j+1)}{2}}(D-(2j+1))!!}
\,\sum_{0\leq i_1<i_2<\cdots <i_{2j}\leq n-2}
\hspace{-5mm}
(L_{i_1}L_{i_2}\cdots L_{i_{2j}}L_{n-1})^{-1}\nonumber\\
&=\sum_{j=1}^{\frac{n}{2}}
\frac{(-1)^{\frac{D}{2}}(-1)^{j-1}M^{D-(2j-1)}
}{2^{\frac{D-2(j-1)}{2}}\pi^{\frac{D-2j}{2}}(D-(2j-1))!!}
\,\sum_{0\leq i_1<i_2<\cdots <i_{2j-2}\leq n-2}
\hspace{-8mm}
(L_{i_1}L_{i_2}\cdots L_{i_{2j-2}}L_{n-1})^{-1}.
\label{newshiki13}
\end{align}
Combining the two results (\ref{newshiki12}) and (\ref{newshiki13}), we obtain 
\begin{align}
F_{L_0,L_1,\cdots, L_{n-1}}^{(n)}\Big|_{\rm n.a.}
=
\sum_{j=1}^{\frac{n}{2}}
\frac{(-1)^{\frac{D}{2}}(-1)^{j-1}M^{D-(2j-1)}
}{2^{\frac{D-2(j-1)}{2}}\pi^{\frac{D-2j}{2}}(D-(2j-1))!!}
\,\sum_{0\leq i_1<i_2<\cdots <i_{2j-1}\leq n-1}
\hspace{-8mm}
(L_{i_1}L_{i_2}\cdots L_{i_{2j-1}})^{-1},
\label{kekka1}
\end{align}
where we have used the relation
\begin{align}
&\hspace{-12mm}
\sum_{0\leq i_1<i_2<\cdots <i_{2j-1}\leq n-2}
\hspace{-8mm}
(L_{i_1}L_{i_2}\cdots L_{i_{2j-1}})^{-1}
+
\sum_{0\leq i_1<i_2<\cdots <i_{2j-2}\leq n-2}
\hspace{-8mm}
(L_{i_1}L_{i_2}\cdots L_{i_{2j-2}}L_{n-1})^{-1}\nonumber\\
&=
\sum_{0\leq i_1<i_2<\cdots <i_{2j-1}\leq n-1}
\hspace{-8mm}
(L_{i_1}L_{i_2}\cdots L_{i_{2j-1}})^{-1}.
\label{newshiki13_2}
\end{align}
%
%
%
%
\subsection{$(D, n)=({\rm even}, {\rm odd})$}
%
%
%
Likewise the previous case, one needs the poles of $\Gamma(t-\frac{\ell}{2})$ for $\ell=$ odd and those of 
$\zeta(2t-\ell)$ for $\ell=$  even in Eq.(\ref{eq2.34}) in order to have the non-analytic terms for $D=$ even. 
Then, we write Eq.(\ref{eq2.34}) as 
\begin{align}
&F_{L_0,L_1,\cdots, L_{n-1}}^{(n)}\Big|_{\rm n.a.}
=-\frac{1}{(2\pi)^{\frac{D}{2}}}
\left(\frac{M^2}{2}\right)^{\frac{D}{2}}
\frac{1}{2\pi i}\int_{c-i\infty}^{c+i\infty}dt~\Gamma(t-\tfrac{D}{2})\left(\frac{ML_{n-1}}{2}\right)^{-2t}\nonumber\\
&\times 
\Biggl(
\sum_{j=1}^{\frac{n-1}{2}}(-1)^{2j-1}\pi^{\frac{2j-1}{2}}
\hspace{-8mm}
\sum_{0\leq i_1<i_2<\cdots <i_{2j-1}\leq n-2}
\hspace{-8mm}
(L_{i_1}L_{i_2}\cdots L_{i_{2j-1}})^{-1}(L_{n-1})^{2j-1}\Gamma(t-\tfrac{2j-1}{2})\zeta(2t-(2j-1))\nonumber\\
&+
\sum_{j=1}^{\frac{n+1}{2}}(-1)^{2j-2}\pi^{\frac{2j-2}{2}}
\hspace{-8mm}
\sum_{0\leq i_1<i_2<\cdots <i_{2j-2}\leq n-2}
\hspace{-8mm}
(L_{i_1}L_{i_2}\cdots L_{i_{2j-2}})^{-1}(L_{n-1})^{2j-2}\Gamma(t-\tfrac{2j-2}{2})\zeta(2t-(2j-2))
\Biggr)\Bigg|_{\rm n.a.}.\nonumber\\
\label{newshiki14}
\end{align}
The residue integral for the pole $t=\frac{2j-1}{2}$of the gamma function in the first term of Eq.(\ref{newshiki14}) yields
\begin{align}
&&-\sum_{j=1}^{\frac{n-1}{2}}
\frac{(-1)^{\frac{D}{2}}(-1)^{j-1}M^{D-(2j-1)}
}{2^{\frac{D-2(j-1)}{2}}\pi^{\frac{D-2j}{2}}(D-(2j-1))!!}
\,\sum_{0\leq i_1<i_2<\cdots <i_{2j-1}\leq n-2}
\hspace{-8mm}
(L_{i_1}L_{i_2}\cdots L_{i_{2j-1}})^{-1},
\label{newshiki14_1}
\end{align}
%
%
and the residue integral for the pole $t=\frac{2j-1}{2}$ of the zeta function 
in the second term in Eq.(\ref{newshiki14}) gives 
\begin{align}
-\sum_{j=1}^{\frac{n+1}{2}}
\frac{(-1)^{\frac{D}{2}}(-1)^{j-1}M^{D-(2j-1)}
}{2^{\frac{D-2(j-1)}{2}}\pi^{\frac{D-2j}{2}}(D-(2j-1))!!}
\,\sum_{0\leq i_1<i_2<\cdots <i_{2j-2}\leq n-2}
\hspace{-8mm}
(L_{i_1}L_{i_2}\cdots L_{i_{2j-2}}L_{n-1})^{-1}.
\label{newshiki15}
\end{align}
We combine the two results (\ref{newshiki14_1}) and (\ref{newshiki15}) to yield
\begin{align}
&F_{L_0,L_1,\cdots, L_{n-1}}^{(n)}\Big|_{\rm n.a.}\nonumber\\
&\hspace{5mm}
=-\sum_{j=1}^{\frac{n+1}{2}}
\frac{(-1)^{\frac{D}{2}}(-1)^{j-1}M^{D-(2j-1)}}{2^{\frac{D-2(j-1)}{2}}\pi^{\frac{D-2j}{2}}(D-(2j-1))!!}\,
\sum_{0\leq i_1<i_2<\cdots <i_{2j-1}\leq n-1}
\hspace{-8mm}
(L_{i_1}L_{i_2}\cdots L_{i_{2j-1}})^{-1}.
\label{kekka2}
\end{align}
%
%
%
%
%
%
\subsection{$(D, n)=({\rm odd}, {\rm even})$}
%
%
%
The space-time dimension $D$ is odd in this case, so that one needs the poles which are integers 
in the integrand of Eq.(\ref{eq2.34}) in order to have the non-analytic terms, that is, the poles 
of $\Gamma(t-\frac{\ell}{2})$ for $\ell=$ even and those of $\zeta(2t-\ell)$ for $\ell=$ odd.
%
%
From Eq.(\ref{eq2.34}), we have the same expression as Eq.(\ref{newshiki11}).
The residue integral for the pole $t=j$ of the zeta function in the first term of Eq.(\ref{newshiki11})
yields
\begin{align}
\sum_{j=1}^{\frac{n}{2}}
\frac{(-1)^{\frac{D-1}{2}}(-1)^jM^{D-2j}}{2^{\frac{D-(2j-1)}{2}}\pi^{\frac{D-(2j+1)}{2}}(D-2j)!!}
\sum_{0\leq i_1<i_2<\cdots <i_{2j-1}\leq n-2}
(L_{i_1}L_{i_2}\cdots L_{i_{2j-1}}L_{n-1})^{-1}
\label{newshiki17}.
\end{align}
For the second term in Eq.(\ref{newshiki11}), the pole $t=j$ of the gamma function gives us
\begin{align}
\sum_{j=0}^{\frac{n}{2}-1}
\frac{(-1)^{\frac{D-1}{2}}(-1)^{j}M^{D-2j}}{2^{\frac{D-(2j-1)}{2}}\pi^{\frac{D-(2j+1)}{2}}(D-2j)!!}
\sum_{0\leq i_1<i_2<\cdots <i_{2j}\leq n-2}
(L_{i_1}L_{i_2}\cdots L_{i_{2j}})^{-1}.
\label{newshiki19}
\end{align}
We put the two results (\ref{newshiki17}) and (\ref{newshiki19}) together to yield
\begin{align}
\hspace{-3mm}
F_{L_0,L_1,\cdots, L_{n-1}}^{(n)}\Big|_{\rm n.a.}
=
\sum_{j=0}^{\frac{n}{2}}
\frac{(-1)^{\frac{D-1}{2}}(-1)^{j}M^{D-2j}}{2^{\frac{D-(2j-1)}{2}}\pi^{\frac{D-(2j+1)}{2}}(D-2j)!!}
\sum_{0\leq i_1<i_2<\cdots <i_{2j}\leq n-1}
(L_{i_1}L_{i_2}\cdots L_{i_{2j}})^{-1}.
\label{kekka3}
\end{align}
Let us note that the $j=0$ contribution does not have the dependence on the scale of the $S^1$, but does the mass scale $M$ alone.
%
%
%
%
\subsection{$(D, n)=({\rm odd}, {\rm odd})$}
%
%
Similar to the previous case, one needs the poles of $\Gamma(t-\frac{\ell}{2})$ for $\ell=$ even and those of 
$\zeta(2t-\ell)$ for $\ell=$ odd in Eq.(\ref{eq2.34}) in order to have the non-analytic terms for $D=$ odd. 
Then, we write Eq.(\ref{eq2.34}) as
\begin{align}
&F_{L_0,L_1,\cdots, L_{n-1}}^{(n)}\Big|_{\rm n.a.}
=-\frac{1}{(2\pi)^{\frac{D}{2}}}
\left(\frac{M^2}{2}\right)^{\frac{D}{2}}
\frac{1}{2\pi i}\int_{c-i\infty}^{c+i\infty}dt~\Gamma(t-\tfrac{D}{2})\left(\frac{ML_{n-1}}{2}\right)^{-2t}
\nonumber\\
&\times
\Biggl(
\sum_{j=1}^{\frac{n-1}{2}}(-1)^{2j-1}\pi^{\frac{2j-1}{2}}
\hspace{-8mm}
\sum_{0\leq i_1<i_2<\cdots <i_{2j-1}\leq n-2}
\hspace{-8mm}
(L_{i_1}L_{i_2}\cdots L_{i_{2j-1}})^{-1}(L_{n-1})^{2j-1}\Gamma(t-\tfrac{2j-1}{2})\zeta(2t-(2j-1))\nonumber\\
&\hspace{8mm}
+
\sum_{j=0}^{\frac{n-1}{2}}(-1)^{2j}\pi^{j}
\hspace{-5mm}
\sum_{0\leq i_1<i_2<\cdots <i_{2j}\leq n-2}
\hspace{-5mm}
(L_{i_1}L_{i_2}\cdots L_{i_{2j}})^{-1}(L_{n-1})^{2j}~\Gamma(t-j)\zeta(2t-2j)
\Biggr)\Bigg|_{\rm n.a.}.\label{newshiki21}
\end{align}
The pole $t=j$ of the zeta function in the the first term of Eq.(\ref{newshiki21}) gives
\begin{align}
-\sum_{j=1}^{\frac{n-1}{2}}
\frac{(-1)^{\frac{D-1}{2}}(-1)^{j}M^{D-2j}}{2^{\frac{D-(2j-1)}{2}}\pi^{\frac{D-(2j+1)}{2}}(D-2j)!!}
\sum_{0\leq i_1<i_2<\cdots <i_{2j-1}\leq n-2}
(L_{i_1}L_{i_2}\cdots L_{i_{2j-1}}L_{n-1})^{-1}.
\label{newshiki22}
\end{align}
The residue integral for the pole $t=j$ of the gamma function in the second 
term of Eq.(\ref{newshiki21}) becomes
\begin{align}
-\sum_{j=0}^{\frac{n-1}{2}}
\frac{(-1)^{\frac{D-1}{2}}(-1)^{j}M^{D-2j}}{2^{\frac{D-(2j-1)}{2}}\pi^{\frac{D-(2j+1)}{2}}(D-2j)!!}
\sum_{0\leq i_1<i_2<\cdots <i_{2j}\leq n-2}
(L_{i_1}L_{i_2}\cdots L_{i_{2j}})^{-1}.
\label{newshiki23}
\end{align}
We combine the two results (\ref{newshiki22}) and (\ref{newshiki23}) to yield
\begin{align}
F_{L_0,L_1,\cdots, L_{n-1}}^{(n)}\Big|_{\rm n.a.}
=
-\sum_{j=0}^{\frac{n-1}{2}}
\frac{(-1)^{\frac{D-1}{2}}(-1)^{j}M^{D-2j}}{2^{\frac{D-(2j-1)}{2}}\pi^{\frac{D-(2j+1)}{2}}(D-2j)!!}
\sum_{0\leq i_1<i_2<\cdots <i_{2j}\leq n-1}
(L_{i_1}L_{i_2}\cdots L_{i_{2j}})^{-1}.\nonumber\\
\label{kekka4}
\end{align}
Let us note again that the $j=0$ contribution 
does not possess the scale dependence on the $S^1$, but does the mass scale $M$.

We have succeeded in obtaining the non-analytic terms in $F_{L_0,L_1,\cdots, L_{n-1}}^{(n)}$
as Eqs.(\ref{kekka1}), (\ref{kekka2}), (\ref{kekka3}) and (\ref{kekka4}), depending 
on even/odd $D$ and $n$. 
In order to calculate the
non-analytic terms in Eq.(\ref{shikinewTneq0}), 
it is useful to generalize the expressions of 
$F_{L_0,L_1,\cdots, L_{n-1}}^{(n)}\Big|_{\rm n.a.}$
to
$F^{(n)}_{L_{i_1},\cdots, L_{i_{n}}}\Big|_{\rm n.a.}$.
Then, we have
\begin{align}
(D, n)=({\rm even},{\rm even}):~F^{(n)}_{L_{i_1},\cdots, L_{i_{n}}}\Big|_{\rm n.a.}&=
\sum_{j=1}^{\frac{n}{2}} 
\frac{(-1)^{\frac{D}{2}+j-1}M^{D-(2j-1)}}{2^{\frac{D-2(j-1)}{2}}\pi^{\frac{D-2j}{2}}(D-(2j-1))!!}\nonumber\\
&\hspace{5mm}
\times \Bigl\{
\sum_{\setTT{\ell_1}{\ell_2}{\ell_{2j-1}}{i_1}{i_2}{i_{n}}}
(L_{\ell_1}L_{\ell_2}\cdots L_{\ell_{2j-1}})^{-1}\Bigr\},
\label{shiki303}\\
(D, n)=({\rm even},{\rm odd}):~F^{(n)}_{L_{i_1},\cdots, L_{i_{n}}}\Big|_{\rm n.a.}&=
\sum_{j=1}^{\frac{n+1}{2}} 
\frac{(-1)^{\frac{D}{2}+j}M^{D-(2j-1)}}{2^{\frac{D-2(j-1)}{2}}\pi^{\frac{D-2j}{2}}(D-(2j-1))!!}
\nonumber\\
&\hspace{5mm}
\times 
\Bigl\{ 
\sum_{\setTT{\ell_1}{\ell_2}{\ell_{2j-1}}{i_1}{i_2}{i_{n}}}
(L_{\ell_1}L_{\ell_2}\cdots L_{\ell_{2j-1}})^{-1}\Bigr\},
\label{shiki304}\\
(D, n)=({\rm odd},{\rm even}):~F^{(n)}_{L_{i_1},\cdots, L_{i_{n}}}\Big|_{\rm n.a.}&=
\sum_{j=0}^{\frac{n}{2}} 
\frac{(-1)^{\frac{D-1}{2}+j}M^{D-2j}}{2^{\frac{D-(2j-1)}{2}}\pi^{\frac{D-(2j+1)}{2}}(D-2j)!!}\nonumber\\
&\hspace{5mm}
\times 
\Bigl\{
\sum_{\setTT{\ell_1}{\ell_2}{\ell_{2j}}{i_1}{i_2}{i_{n}}}
(L_{\ell_1}L_{\ell_2}\cdots L_{\ell_{2j}})^{-1}
\Bigr\},
\label{shiki305}\\
(D, n)=({\rm odd},{\rm odd}):~F^{(n)}_{L_{i_1},\cdots, L_{i_{n}}}\Big|_{\rm n.a.}&=
\sum_{j=0}^{\frac{n-1}{2}}
\frac{(-1)^{\frac{D-1}{2}+j-1}M^{D-2j}}{2^{\frac{D-(2j-1)}{2}}\pi^{\frac{D-(2j+1)}{2}}(D-2j)!!}\nonumber\\
&\hspace{5mm}
\times 
\Bigl\{
\sum_{\setTT{\ell_1}{\ell_2}{\ell_{2j}}{i_1}{i_2}{i_{n}}}
(L_{\ell_1}L_{\ell_2}\cdots L_{\ell_{2j}})^{-1}\Bigr\},
\label{shiki306}
\end{align}
where the order of the elements in the set should be understood to be
$\ell_1<\ell_2 <\cdots <\ell_{2j-1}$ and $i_1<i_2<\cdots <i_{n-1}$, etc. Eqs.(\ref{shiki303}) $\sim$ (\ref{shiki306})
will be used in the next section.
%
%
%
%
\section{Non-analytic terms in effective potential}
%
%
%
Equipped with Eqs.(\ref{shiki303}) $\sim $ (\ref{shiki306}), let us calculate the 
non-analytic terms in the effective potential (\ref{shikinew1_1}). 
We omit the abbreviation ``{\rm n.a.}'' used in the sections $2$ and $3$, which stands for the non-analytic terms. 
One should keep in mind that we are treating the non-analytic terms in this section. 
%
%
%
We introduce
\begin{align}
A^D_j&\equiv \frac{(-1)^{\frac{D}{2}+j-1}M^{D-(2j-1)}}{2^{\frac{D-2(j-1)}{2}}\pi^{\frac{D-2j}{2}}(D-(2j-1))!!}\,,
\label{shiki401}
\\
B^D_j&\equiv \frac{(-1)^{\frac{D-1}{2}+j-1}M^{D-2j}}{2^{\frac{D-(2j-1)}{2}}\pi^{\frac{D-(2j+1)}{2}}(D-2j)!!}
\label{shiki402}
\end{align}
in Eqs.(\ref{shiki303}) $\sim$ (\ref{shiki306}) in order to write them in compact forms.
%
%
%
%
\subsection{$(D, p+1) =$ (even, even)}
%
%
Let us write 
the $F^{(n)}\ (n \ge 1)$ part of the effective potential in Eq.(\ref{shikinew1_1}) 
for $(D, p+1) =$ (even, even) as
\begin{align}
\sum_{n=1}^{p+1} F^{(n)}&=
\sum_{k=1}^{\frac{p+1}{2}}\Bigl\{F^{(2k-1)}+F^{(2k)}\Bigr\}\nonumber\\
&=\sum_{k=1}^{\frac{p+1}{2}}\bigg\{
\sum_{0\leq i_1 <i_2<\cdots <i_{2k-1}\leq p}
F_{L_{i_1}, L_{i_2},\cdots, L_{i_{2k-1}}}^{(2k-1)} +
\sum_{0\leq i_1 <i_2<\cdots <i_{2k}\leq p}
F_{L_{i_1}, L_{i_2},\cdots, L_{i_{2k}}}^{(2k)}
\bigg\}\nonumber\\
&=
\sum_{k=1}^{\frac{p+1}{2}}\bigg\{
\sum_{0\leq i_1 <i_2<\cdots <i_{2k-1}\leq p}~
\sum_{j=1}^{k}(-1)A^D_j
\sum_{
\substack{
\{\ell_1, \ell_2, \cdots ,\ell_{2j-1}\}
\\
\in 
\{i_1,i_2,\cdots,i_{2k-1}\}
}
}
(L_{\ell_1}L_{\ell_2}\cdots L_{\ell_{2j-1}})^{-1}\nonumber\\
&\hspace{5mm}
+
\sum_{0\leq i_1 <i_2<\cdots <i_{2k}\leq p}~
\sum_{j=1}^{k}
A^D_j
\sum_{\substack{
\{\ell_1, \ell_2, \cdots ,\ell_{2j-1}\}\\
\in 
\{i_1,i_2,\cdots,i_{2k}\}
}}
(L_{\ell_1}L_{\ell_2}\cdots L_{\ell_{2j-1}})^{-1}
\bigg\},
\end{align}
where we have used Eqs.(\ref{shikinewTneq0}), (\ref{shiki303}) and (\ref{shiki304}). 
By changing the order of the summations with respect to $j$ and $k$, we obtain 
\begin{align}
\sum_{n=1}^{p+1} F^{(n)}&=
\sum_{j=1}^{\frac{p+1}{2}}\sum_{k=j}^{\frac{p+1}{2}}
\bigg\{
-\sum_{0\leq i_1 <i_2<\cdots <i_{2k-1}\leq p}~
\sum_{
\setTT{\ell_1}{\ell_2}{\ell_{2j-1}}{i_1}{i_2}{i_{2k-1}}
}
+
\sum_{0\leq i_1 <i_2<\cdots <i_{2k}\leq p}~
\sum_{
\setTT{\ell_1}{\ell_2}{\ell_{2j-1}}{i_1}{i_2}{i_{2k}}
}
\bigg\}\nonumber\\
&\hspace{5mm}
\times A^D_j(L_{\ell_1}L_{\ell_2}\cdots L_{\ell_{2j-1}})^{-1}  ,
\label{shiki403}
\end{align}
where we have used the formula
%
\begin{align}
\sum_{k=1}^{\frac{p+1}{2}} \sum_{j=1}^{k}
 = \sum_{j=1}^{\frac{p+1}{2}}\sum_{k=j}^{\frac{p+1}{2}}\,.
\label{add4.4.1}
\end{align}
%

There are ${}_{p+1}C_{2j-1}$ numbers of the independent configurations for $(L_{\ell_1}L_{\ell_2}\cdots L_{\ell_{2j-1}})^{-1}$ for 
a given value of $j$, each of which has the same multiplicity in Eq.(\ref{shiki403}) 
for a fixed value of $k$.  Hence, it is 
enough to calculate the multiplicity for the configuration 
with $\ell_1=0, \ell_2=1,\cdots, \ell_{2j-1}=2j-2$, that is, $(L_0L_1\cdots L_{2j-2})^{-1}$ 
for each value of $k=j, j+1,\cdots, \frac{p+1}{2}$.

Let us first compute the multiplicity for the case of $k=j$ with fixed $j$. 
Aside form the factor $A_j^D$, we have 
\begin{align}
&\bigg\{
-\sum_{0\leq i_1 <i_2<\cdots <i_{2j-1}\leq p}~
\sum_{\setTT{0}{1}{2j-2}{i_1}{i_2}{i_{2j-1}}}
+
\sum_{0\leq i_1 <i_2<\cdots <i_{2j}\leq p}~
\sum_{\setTT{0}{1}{2j-2}{i_1}{i_2}{i_{2j}}}
\bigg\}
(L_0L_1\cdots L_{2j-2})^{-1} \nonumber\\
&\hspace{5mm}
=
\Bigl\{-1+{}_{p+1-(2j-1)}C_1\Bigr\}(L_0L_1\cdots L_{2j-2})^{-1}.
\end{align}
Let us next calculate the multiplicity for the case of  $k=j+1$. 
The result is given by
\begin{align}
&\bigg\{
-\sum_{0\leq i_1 <i_2<\cdots <i_{2j+1}\leq p}~~
\sum_{\substack{
\{0,1,\cdots, 2j-2\}\\
\in 
\{i_1, i_2, \cdots, i_{2j+1}\}}}
+
\sum_{0\leq i_1 <i_2<\cdots <i_{2j+2}\leq p}~~
\sum_{\substack{\{0,1,\cdots, 2j-2\}\\
\in 
\{i_1, i_2,\cdots, i_{2j+2}\}
}}
\bigg\}\nonumber\\
&\hspace{5mm}\times 
(L_0L_1\cdots L_{2j-2})^{-1} \nonumber\\
&=
\Bigl\{-~{}_{p+1-(2j-1)}C_2+{}_{p+1-(2j-1)}C_3\Bigr\}(L_0L_1\cdots L_{2j-2})^{-1}.
\end{align}
In the same manner, one can calculate the multiplicity for $k=j+2, j+3,\cdots ,\frac{p+1}{2}$.
Collecting the terms with $k=j, j+1, \cdots, \frac{p+1}{2}$ and 
denoting $A\equiv p+1-(2j-1)$, which is odd for the 
present case, we have
\begin{align}
&\Big\{
-\bigl(1+{}_AC_2 +{}_AC_4+\cdots +{}_AC_{A-1}\bigr)
+\bigl({}_AC_1+{}_AC_3+\cdots +{}_AC_A\bigr)
\Bigr\}(L_0L_1\cdots L_{2j-2})^{-1}\nonumber\\
&=
\Big\{-\half \times 2^A +\half \times 2^A
\Bigr\}(L_0L_1\cdots L_{2j-2})^{-1}=0.
\label{shiki404}
\end{align}
This holds for any $j$ satisfying $1\leq j\leq \frac{p+1}{2}$ and the same 
conclusion (\ref{shiki404}) holds for the other configurations
of $L_{\ell_{1}}L_{\ell_{2}}\cdots L_{\ell_{2j-1}}$.
Furthermore, Eq.(\ref{shiki2102}) is analytic for $D=$ even.
Thus, we conclude that $V_{\rm eff}\Big|_{\rm n.a.}=0$, that is, there is no non-analytic term in the 
effective potential for $D=$ even and $p+1=$ even.
%
%
%
%
\subsection{$(D, p+1)=$ (even, odd)}
%
%
We write 
the $F^{(n)}\ (n \ge 1)$ part of the effective potential for this case as
\begin{align}
\sum_{n=1}^{p+1} F^{(n)}&=
\sum_{k=1}^{\frac{p}{2}}\Bigl\{F^{(2k-1)}+F^{(2k)}\Bigr\}+F^{(p+1)}
\nonumber\\
&=
\sum_{k=1}^{\frac{p}{2}}\bigg\{
\sum_{0\leq i_1 <i_2<\cdots <i_{2k-1}\leq p}~
\sum_{j=1}^{k}(-1)A^D_j
\sum_{\setTT{\ell_1}{\ell_2}{\ell_{2j-1}}{i_1}{i_2}{i_{2k-1}}
}
(L_{\ell_1}L_{\ell_2}\cdots L_{\ell_{2j-1}})^{-1}\nonumber\\
&\hspace{5mm}
+
\sum_{0\leq i_1 <i_2<\cdots <i_{2k}\leq p}~
\sum_{j=1}^{k}A^D_j
\sum_{
\setTT{\ell_1}{\ell_2}{\ell_{2j-1}}{i_1}{i_2}{i_{2k}}
}
(L_{\ell_1}L_{\ell_2}\cdots L_{\ell_{2j-1}})^{-1}
\bigg\}\nonumber\\
&\hspace{5mm}
+
\sum_{j=1}^{\frac{p+2}{2}}
(-1)A^D_j\sum_{0\leq \ell_1<\ell_2<\cdots <\ell_{2j-1}\leq p}
(L_{\ell_1}L_{\ell_2}\cdots L_{\ell_{2j-1}})^{-1},
\label{shiki405}
\end{align}
where we have again used Eqs.(\ref{shikinewTneq0}), (\ref{shiki303}) and (\ref{shiki304}). 
The last term in Eq.(\ref{shiki405})
with $j=\frac{p+2}{2}$ yields $-A^D_{\frac{p+2}{2}}(L_0L_1\cdots L_{p})^{-1}$. 
Then, we have
\begin{align}
\sum_{n=1}^{p+1} F^{(n)}&=
\sum_{j=1}^{\frac{p}{2}}\Biggl(\sum_{k=j}^{\frac{p}{2}}
\bigg\{
-\sum_{0\leq i_1 <i_2<\cdots <i_{2k-1}\leq p}~
\sum_{
\substack{
\{\ell_1,\ell_2,\cdots,\ell_{2j-1}\}\\
\in
\{i_1,i_2,\cdots, i_{2k-1}\}
}}
+
\sum_{0\leq i_1 <i_2<\cdots <i_{2k}\leq p}~
\sum_{
\substack{
\{\ell_1,\ell_2,\cdots,\ell_{2j-1}\}\\
\in
\{i_1,i_2,\cdots, i_{2k}\}
}}
\bigg\}\nonumber\\
&\hspace{5mm}
-\sum_{0\leq \ell_1<\ell_2<\cdots <\ell_{2j-1}\leq p}\,
\Biggr)
A^D_j(L_{\ell_1}L_{\ell_2}\cdots L_{\ell_{2j-1}})^{-1} 
-A^D_{\frac{p+2}{2}}(L_0L_1\cdots L_{p})^{-1} ,
\label{shiki406}
\end{align}
where we have exchanged the order of the summations with respect to $j$ and $k$.

Let us count the multiplicity for the configuration with 
$\ell_1=0,\ell_2=1,\cdots, \ell_{2j-1}=2j-2$.
Note that the third term in Eq.(\ref{shiki406}) results one for the configuration. 
We calculate the multiplicity 
of the first and the second terms in Eq.(\ref{shiki406}) for $k=j$. Apart from the 
factor $A_j^D$, it is calculated as
\begin{align}
&\bigg\{
-\sum_{0\leq i_1 <i_2<\cdots <i_{2j-1}\leq p}~
\sum_{
\substack{
\{0,1,\cdots, 2j-2\}\\
\in
\{i_1,i_2,\cdots, i_{2j-1}\}
}}
+
\sum_{0\leq i_1 <i_2<\cdots <i_{2j}\leq p}~
\sum_{
\substack{
\{0,1,\cdots, 2j-2\}\\
\in
\{i_1,i_2,\cdots, i_{2j}\}
}}
\bigg\}
(L_0L_1\cdots L_{2j-2})^{-1}
\nonumber\\
&
=
\Bigl\{-1+{}_{p+1-(2j-1)}C_1 \Bigr\}(L_0L_1\cdots L_{2j-2})^{-1}.
\end{align}
Likewise, for $k=j+1$, we have
\begin{align}
&\bigg\{
-\sum_{0\leq i_1 <i_2<\cdots <i_{2j+1}\leq p}~
\sum_{
\substack{
\{0,1,\cdots, 2j-2\}\\
\in
\{i_1,i_2,\cdots, i_{2j+1}\}
}}
+
\sum_{0\leq i_1 <i_2<\cdots <i_{2j+2}\leq p}~
\sum_{
\substack{
\{0,1,\cdots, 2j-2\}\\
\in
\{i_1,i_2,\cdots, i_{2j+2}\}
}}
\bigg\}
(L_0L_1\cdots L_{2j-2})^{-1}\nonumber\\
&=
\Bigl\{-~{}_{p+1-(2j-1)}C_{2} +{}_{p+1-(2j-1)}C_3\Bigr\}(L_0L_1\cdots L_{2j-2})^{-1}.
\end{align}
Similarly, we can compute the multiplicity for $k=j+1, j+2, \cdots, \frac{p}{2}$.
Collecting all the terms with $k=j, j+1, \cdots, \frac{p}{2}$ and including 
the third term in Eq.(\ref{shiki406}), we have
\begin{align}
&\Big\{
-\bigl(1+{}_AC_2 +{}_AC_4+\cdots +{}_AC_{A-2}\bigr)
+\bigl({}_AC_1+{}_AC_3+\cdots +{}_AC_{A-1}\bigr)-1
\Bigr\}(L_0L_1\cdots L_{2j-2})^{-1}\nonumber\\
&=
\Big\{-\Big(\half \times 2^A -{}_AC_A\Big)+\half \times 2^A -1
\Bigr\}(L_0L_1\cdots L_{2j-2})^{-1}=0\,,
\end{align}
where $A\equiv p+1-(2j-1)$, which is even in this case.
This holds for any value of $j$ between $1$ and $\frac{p}{2}$.  
Hence, the last term in Eq.(\ref{shiki406}) alone is left to yield the non-analytic term in the effective potential 
for $D=$ even and $p+1=$ odd, i.e.
\begin{align}
V_{\rm eff}\Big|_{\rm n.a.}
&=-A^D_{\frac{p+2}{2}}(L_0L_1\cdots L_{p})^{-1} \nonumber\\
&=\frac{(-1)^{\frac{D+p+2}{2}}}{2^{\frac{D-p}{2}}\pi^{\frac{D-(p+2)}{2}}(D-(p+1))!!}~
\frac{M^{D-(p+1)}}{L_0L_1\cdots L_{p}} .
\label{shikiresult1}
\end{align}
%
%
%
\subsection{$(D, p+1)=$ (odd, even)}
%
%
The $F^{(n)}\ (n \ge 1)$ part of the effective potential for this case is given by
\begin{align}
\sum_{n=1}^{p+1} F^{(n)}&=
\sum_{k=1}^{\frac{p+1}{2}}\Bigl\{F^{(2k-1)}+F^{(2k)}\Bigr\}\nonumber\\
&=
\sum_{k=1}^{\frac{p+1}{2}}\bigg\{
\sum_{0\leq i_1 <i_2<\cdots <i_{2k-1}\leq p}~
\sum_{j=0}^{k-1}
B^D_j
\sum_{\setTT{\ell_1}{\ell_2}{\ell_{2j}}{i_1}{i_2}{i_{2k-1}}
}
(L_{\ell_1}L_{\ell_2}\cdots L_{\ell_{2j}})^{-1}\nonumber\\
&\hspace{5mm}
+
\sum_{0\leq i_1 <i_2<\cdots <i_{2k}\leq p}~
\sum_{j=0}^{k}
(-1)B^D_j
\sum_{\setTT{\ell_1}{\ell_2}{\ell_{2j}}{i_1}{i_2}{i_{2k}}
}
(L_{\ell_1}L_{\ell_2}\cdots L_{\ell_{2j}})^{-1}
\bigg\},
\label{shiki407}
\end{align}
where we have used Eqs.(\ref{shikinewTneq0}), (\ref{shiki305}) and (\ref{shiki306}). 
Separating the $j=k$ contribution from the second term 
in Eq.(\ref{shiki407}) and exchanging the summations with respect to $j$ and $k$,
we obtain
\begin{align}
\sum_{n=1}^{p+1} F^{(n)}&=
\sum_{j=0}^{\frac{p-1}{2}}\sum_{k=j+1}^{\frac{p+1}{2}}
\bigg\{
\sum_{0\leq i_1 <i_2<\cdots <i_{2k-1}\leq p}~
\sum_{
\substack{\{\ell_1,\ell_2,\cdots, \ell_{2j}\}\\
\in
\{i_1,i_2,\cdots, i_{2k-1}\}
}
}-
\sum_{0\leq i_1 <i_2<\cdots <i_{2k}\leq p}~
\sum_{
\substack{
\{\ell_1,\ell_2,\cdots, \ell_{2j}\}\\
\in
\{i_1,i_2,\cdots, i_{2k}\}
}
}
\bigg\} \nonumber\\
&\hspace{5mm}
\times B^D_j(L_{\ell_1}L_{\ell_2}\cdots L_{\ell_{2j}})^{-1} \nonumber\\
&\hspace{5mm}
-\sum_{k=1}^{\frac{p-1}{2}}~\sum_{0\leq \ell_1<\ell_2<\cdots <\ell_{2k}\leq p}
B^D_k(L_{\ell_1}L_{\ell_2}\cdots L_{\ell_{2k}})^{-1}  
-B^D_{\frac{p+1}{2}}(L_0L_1\cdots L_{p})^{-1}.
\label{shiki409}
\end{align}
Here, the last term in Eq.(\ref{shiki409}) has been separated from the
third term and corresponds to the $k=\frac{p+1}{2}$ contribution of it.

Let us first study the $j=0$ contribution in the first and the second term in Eq.(\ref{shiki409}), which is given by
\begin{align}
&\sum_{k=1}^{\frac{p+1}{2}
}\Bigl\{{}_{p+1}C_{2k-1}-{}_{p+1}C_{2k}
\Bigr\}B_0^D\nonumber\\
%
%
&=\Bigl\{\half \times 2^{p+1}-\Big(\half\times 2^{p+1} -{}_{p+1}C_0\Big)\Bigr\}B_0^D 
=-\frac{(-1)^{\frac{D-1}{2}}}{2^{\frac{D+1}{2}}\pi^{\frac{D-1}{2}}D!!}M^D.
\end{align}
This exactly cancels the contribution coming from the one-loop correction
for the zero modes (\ref{shiki2102}) for $D=$ odd.\ 
Then, rewriting $k$ as $j$ in the third term in Eq. (\ref{shiki409}), we have
\begin{align}
F^{(0)}+\sum_{n=1}^{p+1} F^{(n)}&=
\sum_{j=1}^{\frac{p-1}{2}}\Biggl(
\sum_{k=j+1}^{\frac{p+1}{2}}
\bigg\{
\sum_{0\leq i_1 <i_2<\cdots <i_{2k-1}\leq p}~
\sum_{
\substack{\{\ell_1,\ell_2,\cdots, \ell_{2j}\}\\
\in
\{i_1,i_2,\cdots, i_{2k-1}\}
}
}
-\sum_{0\leq i_1 <i_2<\cdots <i_{2k}\leq p}~
\sum_{
\substack{
\{\ell_1,\ell_2,\cdots, \ell_{2j}\}\\
\in
\{i_1,i_2,\cdots, i_{2k}\}
}
}
\bigg\}  \nonumber\\
&\hspace{5mm}
-
\sum_{0\leq \ell_1<\ell_2<\cdots <\ell_{2j}\leq p}
\Biggr)B^D_j(L_{\ell_1}L_{\ell_2}\cdots L_{\ell_{2j}})^{-1}
-B^D_{\frac{p+1}{2}}(L_0L_1\cdots L_{p})^{-1}.
\label{shiki410}
\end{align}

Let us count the multiplicity for the configuration 
$\ell_1=0, \ell_2=1,\cdots, \ell_{2j}=2j-1$
for fixed $j$. To this end, we extract all the terms proportional to $L_{0}L_{1}\cdots L_{2j-1}$
for $k=j+1, j+2, \cdots, \frac{p+1}{2}$ from Eq.(\ref{shiki410}).
Apart from the factor $B_{j}^{D}$, the result is
\begin{align}
&\Big\{
\bigl({}_BC_1 +{}_BC_3+\cdots +{}_BC_{B-1}\bigr)
-\bigl({}_BC_2+{}_BC_4+\cdots +{}_BC_B\bigr)-1
\Bigr\}(L_0L_1\cdots L_{2j-1})^{-1}\nonumber\\
&=
\Big\{\Big(\half \times 2^B\Big) -\Big(\half \times 2^B -{}_BC_0\Big)-1
\Bigr\}(L_0L_1\cdots L_{2j-1})^{-1}=0,
\label{add4.26}
\end{align}
where $B=p+1-2j$, which is even in this case.
Since Eq.(\ref{add4.26}) holds by any values of $j=1,2,\cdots, \frac{p-1}{2}$,
only the last term in Eq.(\ref{shiki410}) is left to yield the non-analytic term in the effective potential,
\begin{align}
V_{\rm eff}\Big|_{\rm n.a.}&=-B^D_{\frac{p+1}{2}}(L_0L_1\cdots L_{p})^{-1} \nonumber\\
&=\frac{(-1)^{\frac{D+p}{2}}}{2^{\frac{D-p}{2}}\pi^{\frac{D-(p+2)}{2}}(D-(p+1))!!}~
\frac{M^{D-(p+1)}}{L_0L_1\cdots L_{p}} .
\label{shikiresult2}
\end{align}
%
%
%
\subsection{$(D, p+1)=$ (odd, odd)}
%
%
%
Let us proceed the fourth case.
The $F^{(n)}\ (n \ge 1)$ part of the effective potential is given by
\begin{align}
\sum_{n=1}^{p+1} F^{(n)}&=
\sum_{k=1}^{\frac{p}{2}}\Bigl\{F^{(2k-1)}+F^{(2k)}\Bigr\}+F^{(p+1)}\nonumber\\
&=
\sum_{k=1}^{\frac{p}{2}}\bigg\{
\sum_{0\leq i_1 <i_2<\cdots <i_{2k-1}\leq p}~
\sum_{j=0}^{k-1}
B^D_j
\sum_{\setTT{\ell_1}{\ell_2}{\ell_{2j}}{i_1}{i_2}{i_{2k-1}}}
(L_{\ell_1}L_{\ell_2}\cdots L_{\ell_{2j}})^{-1}\nonumber\\
&\hspace{5mm}
+
\sum_{0\leq i_1 <i_2<\cdots <i_{2k}\leq p}~
\sum_{j=0}^{k}
(-1)B^D_j
\sum_{\setTT{\ell_1}{\ell_2}{\ell_{2j}}{i_1}{i_2}{i_{2k}}}
(L_{\ell_1}L_{\ell_2}\cdots L_{\ell_{2j}})^{-1}
\bigg\}\nonumber\\
&\hspace{5mm}
+
\sum_{j=0}^{\frac{p}{2}}B_j^D
\sum_{0\leq \ell_1<\ell_2<\cdots <\ell_{2j}\leq p}
(L_{\ell_1}L_{\ell_2}\cdots L_{\ell_{2j}})^{-1},
\end{align}
where we have again used Eqs.(\ref{shikinewTneq0}), (\ref{shiki305}) and (\ref{shiki306}). 
We write the above equation as
\begin{align}
\sum_{n=1}^{p+1} F^{(n)}&=
\sum_{k=1}^{\frac{p}{2}}
\sum_{j=0}^{k-1}
\bigg\{
\sum_{0\leq i_1 <i_2<\cdots <i_{2k-1}\leq p}
B^D_j
\sum_{\setTT{\ell_1}{\ell_2}{\ell_{2j}}{i_1}{i_2}{i_{2k-1}}
}
(L_{\ell_1}L_{\ell_2}\cdots L_{\ell_{2j}})^{-1}\nonumber\\
&\hspace{5mm}
+
\sum_{0\leq i_1 <i_2<\cdots <i_{2k}\leq p}
(-1)B^D_j
\sum_{\setTT{\ell_1}{\ell_2}{\ell_{2j}}{i_1}{i_2}{i_{2k}}
}
(L_{\ell_1}L_{\ell_2}\cdots L_{\ell_{2j}})^{-1}
\bigg\}\nonumber\\
&\hspace{5mm}
+
\sum_{k=1}^{\frac{p}{2}}~\sum_{0\leq i_1 <i_2<\cdots <i_{2k}\leq p}
\hspace{-5mm}
(-1)B^D_k(L_{i_1}L_{i_2}\cdots L_{i_{2k}})^{-1}
+
\sum_{j=0}^{\frac{p}{2}}B_j^D
\sum_{0\leq \ell_1<\ell_2<\cdots <\ell_{2j}\leq p}
\hspace{-5mm}
(L_{\ell_1}L_{\ell_2}\cdots L_{\ell_{2j}})^{-1} .\nonumber\\
\label{shiki411}
\end{align}
Note that the third term in Eq.(\ref{shiki411}) cancels the last one in Eq.(\ref{shiki411})
except for the $j=0$ term, which cancels the one-loop contribution 
for the zero modes (\ref{shiki2102}) for $D=$ odd.

Then, what is left is the first and the second terms in Eq.(\ref{shiki411}). 
By exchanging the summations with respect to $j$ and $k$, we obtain 
\begin{align}
F^{(0)}+\sum_{n=1}^{p+1} F^{(n)}&=
\sum_{j=0}^{\frac{p-2}{2}}\sum_{k=j+1}^{\frac{p}{2}}
\bigg\{
\sum_{0\leq i_1 <i_2<\cdots <i_{2k-1}\leq p}~
\sum_{
\substack{
\{\ell_1,\ell_2,\cdots,\ell_{2j}\}
\\\in
\{i_1,i_2,\cdots, i_{2k-1}\}
}}
-\sum_{0\leq i_1 <i_2<\cdots <i_{2k}\leq p}~
\sum_{
\substack{
\{\ell_1,\ell_2,\cdots,\ell_{2j}\}
\\
\in
\{i_1,i_2,\cdots, i_{2k}\}
}}
\bigg\} \nonumber\\
&\hspace{5mm}
\times B^D_j(L_{\ell_1}L_{\ell_2}\cdots L_{\ell_{2j}})^{-1}  .
\label{shiki412}
\end{align}
Let us count the multiplicity for the configuration with 
$\ell_1=0, \ell_2=1,\cdots, \ell_{2j}=2j-1$ 
for fixed $j$ by extracting all the terms proportional to $L_{0}L_{1} \cdots L_{2j-1}$
for $k=j+1, j+1, \cdots, \frac{p}{2}$ from Eq.(\ref{shiki412}).
Apart from the factor $B_{j}^{D}$, the result is given by
\begin{align}
&\hspace{-3mm}
\Big\{
\bigl({}_BC_1 +{}_BC_3+\cdots +{}_BC_{B-2}\bigr)
-\bigl({}_BC_2+{}_BC_4+\cdots +{}_BC_{B-1}\bigr)
\Bigr\}(L_0L_1\cdots L_{2j-1})^{-1}\nonumber\\
&=
\Big\{(\half \times 2^B- {}_BC_B) -(\half \times 2^B -{}_BC_0)
\Bigr\}(L_0L_1\cdots L_{2j-1})^{-1}=0\,,
\end{align}
where $B = p+1-2j$, which is odd in this case.
This holds for any $j$ satisfying $0\leq j\leq \frac{p-2}{2}$, so that
we have $V_{\rm eff}\Big|_{\rm n.a.}=0$, that is, the 
effective potential does not possess non-analytic terms for $D=$ odd and $p+1=$ odd.

We have calculated the non-analytic terms in the effective potential for
any $D$ and $p$. We have found that there is no non-analytic term 
for $(D, p+1) =$ (even, even) and (odd, odd).
On the other hand, the non-analytic term appears for
$(D, p+1) =$ (even, odd) and (odd, even), as shown in 
Eqs.(\ref{shikiresult1}) and (\ref{shikiresult2}), respectively.
The results are summarized as
%
\begin{align}
V_{\rm eff}\Big|_{\rm n.a.}
 = \begin{cases}
     0 & \text{for $(D, p+1) =$ (even, even), (odd, odd)}\,,\\
     \frac{(-1)^{\frac{D+p}{2}}(-1)^{p+1}}{2^{\frac{D-p}{2}}\pi^{\frac{D-(p+2)}{2}}(D-(p+1))!!}~
\frac{M^{D-(p+1)}}{L_0L_1\cdots L_{p}} 
       & \text{for $(D, p+1) =$ (even, odd), (odd, even)}\,.
   \end{cases}
\label{shiki413}
\end{align}
%
The famous Dolan-Jackiw term corresponds to the case of $D=4, p=0$ case in Eq.(\ref{shiki413}). 
We present some results followed from Eq.(\ref{shiki413}) in the table $1$.

We observe that the non-analytic term appears in the effective potential
when $D-(p+1)=$ odd, which corresponds to the odd uncompactified spacial dimensions.
It must be noticed that there is only one non-analytic term  for the given
values of $D$ and $p$, for which the non-analytic term exists.

\begin{table}[ht]
\begin{center}
$S^{1}_{\tau}\times R^{D-(p+1)} \times \prod_{i=1}^{p}S^{1}_{i}$\\
\begin{tabular}{c|c|c|c|c|c}
 & 	$p=0$ &$p=1$&$p=2$&$p=3$&$p=4$ \\\hline 
$D=3$ 	& {\footnotesize non} & $\half \frac{MT}{L_1}$ &{\footnotesize non}   & $\times$ &$\times$\\[0.3cm]
$D=4$ 	& $-\frac{M^3T}{12\pi}$&{\footnotesize non}   &$\half \frac{MT}{L_1L_2}$  &  {\footnotesize non} &$\times$\\[0.3cm]
$D=5$ 	&{\footnotesize non}  &$-\frac{1}{12\pi }\frac{M^3T}{L_1}$ &{\footnotesize non}   & $\half \frac{MT}{L_1L_2L_3}$ &{\footnotesize non} \\[0.3cm]
$D=6$ 	& $\frac{M^5T}{120\pi^2}$ & {\footnotesize non}  & $-\frac{1}{12\pi }\frac{M^3T}{L_1L_2}$  & {\footnotesize non}& $\half \frac{MT}{L_1L_2L_3L_4}$\\[0.3cm]
$D=7$  &{\footnotesize non}  & $\frac{1}{120\pi^2}\frac{M^5T}{L_1}$&{\footnotesize non}  &$-\frac{1}{12\pi }\frac{M^3T}{L_1L_2L_3}$  & {\footnotesize non} 
\\\hline
\end{tabular}
\end{center}
\caption{Some results on non-analytic terms in the effective potential on the space-time 
$S_{\tau}^1\times R^{D-(p+1)}\times \prod_{i=1}^pS_i^1$. 
We use $T (=L_0^{-1})$. 
The ``non'' in the table means
that there is no non-analytic term in the effective potential 
and the ``$\times $'' stands for the case, where the condition $D\geq p+1$ is not satisfied. }
\end{table}
%
%
%
%
%
%
%
%
\section{Conclusions and Discussions}
%
%
We have studied the non-analytic terms in the effective potential for the real scalar field 
at finite temperature in one-loop approximation on the $D$-dimensional space-time, $S_{\tau}^1\times R^{D-(p+1)}\times \prod_{i=1}^pS_i^1$.
The effective potential is given in terms of the modified Bessel function of the second kind accompanied with the multiple
mode summations. We have introduced the integral representation for the modified Bessel function of the second 
kind and have also made use of the analytical extension for the mode summations. The effective potential
is recast into the integral form in the complex plane, and the non-analytic terms are obtained by the residue theorem.

We have clarified the pole structure of the analytical extension for the mode summations
and have found the recurrence relation  (\ref{eq2.31}), from which we have obtained 
the general form (\ref{eq2.32}) for the relevant terms in the mode summations.
We have calculated the non-analytic terms, Eqs.(\ref{shiki303}) $\sim$ (\ref{shiki306}) by the residue 
theorem for the poles of the gamma and zeta functions in Eq.(\ref{eq2.32}) with Eq.(\ref{eq2.33}), depending 
on the even/odd $D$ and $n~(n=1,2,\cdots, p+1)$. The positions of the poles that yield the non-analytic terms are found to 
be independent of $D$. Equipped with them, we have calculated the non-analytic terms in the effective potential, which 
is given by Eq.(\ref{shiki413}), including the famous Dolan-Jackiw term. 
Some explicit results are summarized in the table $1$.

We have found that the effective potential has 
the non-analytic term when the dimension of the flat Euclidean space, $D-(p+1)$ is odd. There is 
only one non-analytic term  for the given values of $D$ and $p$, for which the non-analytic term exists.

There are untouched issues in the paper. We have not discussed the physical origin of such the non-analytic term
in the effective potential. The paper \cite{dj} showed that the famous Dolan-Jackiw term had been emerged 
through the zero mode of the $S_{\tau}^1$ direction, reflecting the infrared dynamics of the theory. It may be important
to clarify the physical origin of the non-analytic term found in this paper. Moreover, it may be 
important to study the physical implication of the non-analytic terms on, for example, the phase transition at
finite temperature.

For the case of the fermion, the boundary condition for the $S_{\tau}^1$ direction
is anti-periodic due to the Fermi statistics, and we have the factor $(-1)^{m_0}$ in the
mode summation. 
Then, the zeta function $\zeta(2t)$ in Eq.(\ref{shiki2106}) is replaced by the eta function, $-\eta(2t)$, which does 
not possess the pole, so that the non-analytic term does not arise for $D=4, p=0$ \cite{dj}. It 
is expected that the analytical extension for the mode summations is modified 
to yield different non-analytic terms from those of the scalar field. Moreover, there are degrees of freedom to 
choose the periodic or anti-periodic boundary condition 
for each spacial $S_i^1~(i=1,2,\cdots, p)$ direction for the fermion (scalar) field.
We also expect that the analytical extension for the mode summations is different from that of
the case for the periodic boundary condition. Accordingly, we may have a different type of non-analytic terms. 
These will be reported elsewhere.

\begin{center}
{\bf Acknowledgement}
\end{center}
This work is supported in part by Grants-in-Aid for Scientific 
Research [No.~18K03649 (M.S.)] from the Ministry of 
Education, Culture, Sports, Science and Technology (MEXT) in Japan.
%
%
\appendix
%
%
%
%
\section{Mode summations and analytical extension}
\label{sec:modenon}
%
%
%
The effective potential contains the modified Bessel function of the second kind 
accompanied with the mode summations. 
In addition to the integal representation (\ref{shiki2105}) in the text 
for the modified Bessel function of the second kind, one needs an analytical extension 
for the mode summations in order to recast the effective potential into the integral form 
and to obtain the non-analytic terms by the residue integral. 
In this appendix, we present an analytical extension for the multiple
mode summations and clarify its pole structure in the residue integral.
%
%
\subsection{Double mode summations}
%
%
The double summations (\ref{shiki2115}) in the text has an analytical extension \cite{elizalde}.
In our analysis, the following formula plays a crucial role:
\begin{align}
&\Gamma(t)\sum_{m_0=1}^{\infty}\{(m_0L_0)^2+c^2\}^{-t}\nonumber\\
&=
-\half  \frac{\Gamma(t)}{c^{2t}} 
+ \frac{\sqrt{\pi}}{2L_0}\frac{\Gamma(t-\half)}{c^{2(t-\half)}}
+\frac{2\pi^t}{L_0^{t+\half}}\frac{1}{c^{t-\half}}
\sum_{n_0=1}^{\infty}n_0^{t-\half}~K_{t-\half}\left(\frac{2\pi n_0}{L_0}c\right)
\nonumber\\
&=
-\half  \frac{\Gamma(t)}{c^{2t}} 
+ \frac{\sqrt{\pi}}{2L_0}\frac{\Gamma(t-\half)}{c^{2(t-\half)}}\nonumber\\
&\hspace{5mm}
+
\frac{1}{\sqrt{\pi}}\left(\frac{\pi}{L_0}\right)^{2t}
\frac{1}{2\pi i}\int_{c_{1} - i\infty}^{c_{1} +i\infty}d{t_{1}}~\Gamma(t_{1} -t +\tfrac{1}{2})
\zeta(2t_{1}-2t +1)\Gamma(t_{1}) 
\left(c\frac{\pi}{L_0}\right)^{-2t_{1}},
\label{appshiki1}
\end{align}
where we have used Eq.(\ref{shiki2105}) in the last equality. 
In obtaining Eq.(\ref{appshiki1}), we have made use of
the Poisson summation
%
\begin{align}
\sum_{m_0=-\infty}^{\infty}\e^{-[(m_0+\eta_{0})L_0]^2t}=\sum_{n_0
=-\infty}^{\infty}\frac{1}{L_0}\left(\frac{\pi}{t}\right)^{\half}\e^{-\frac{(2\pi n_0)^2}{4L_0^2}\frac{1}{t} + 2\pi i n_{0}\eta_{0}}
\label{eqA.2}
\end{align}
%
and the formula
\begin{align}
K_{\nu}(z)=\half\left(\frac{z}{2}\right)^{\nu}\int_0^{\infty}\!dt~t^{-\nu-1}\e^{-\frac{1}{t}(\frac{z}{2})^2-t}.
\label{eqA.3}
\end{align}

Setting $c=m_1L_1$ and taking the summation $\sum_{m_1=1}^{\infty}$, we arrive at
\begin{align}
&\Gamma(t)\sum_{m_0=1}^{\infty}\sum_{m_1=1}^{\infty}\bigl\{(m_0L_0)^2+(m_1 L_1)^2\bigr\}^{-t} \nonumber\\
&= -\half \frac{1}{L_1^{2t}}\Gamma(t)\zeta(2t) 
   + \frac{\sqrt{\pi}}{2}\frac{1}{L_0L_1^{2t-1}}\Gamma(t-\tfrac{1}{2})\zeta(2t -1)\nonumber\\
&\hspace{5mm}
+\frac{1}{\sqrt{\pi}}\left(\frac{\pi}{L_0}\right)^{2t}\frac{1}{2\pi i}
\int_{c_{1} -i\infty}^{c_{1}+i\infty}d{t_{1}}~\Gamma(t_{1} -t +\tfrac{1}{2})
\zeta(2t_{1} -2t +1)\Gamma(t_{1})\zeta(2t_{1})
\left(\pi\frac{L_1}{L_0}\right)^{-2t_{1}} \hspace{-2mm}.
\label{eqA.4}
\end{align}
This is Eq.(\ref{shiki2117_1}) in the text.

As we will show below, an important observation is that the third term in 
Eq.(\ref{eqA.4}) has the property
%
\begin{align}
\frac{1}{2\pi i}\int_{c-i\infty}^{c+i\infty}dt\,
 f(t)\,\frac{1}{2\pi i}\int_{c_{1}-i\infty}^{c_{1}+i\infty}dt_{1}\,
  \Gamma(t_{1}-t+\tfrac{1}{2}) \zeta(2t_{1}-2t+1) \Gamma(t_{1}) \zeta(2t_{1})
   \bigg(\pi\frac{L_{1}}{L_{0}}\bigg)^{-2t_{1}}
    = 0\,,
\label{eqA.5}
\end{align}
%
where $f(t)$ is any function that has no poles inside the region of the
residue integral with respect to $t$.
Eq.(\ref{eqA.5}) implies that the poles coming from the third term of
Eq.(\ref{eqA.4}) do not contribute to the residue integral with
respect to $t$ in Eq.(\ref{eqA.5}).
A similar property holds for the multiple mode summations, as we will see
in the next subsection.

To show Eq.(\ref{eqA.5}), we first note that the combination of 
$\Gamma(t_{1}-\frac{a}{2}) \zeta(2t_{1}-a)$ has no poles except
for the pole of $\Gamma(t_{1}-\frac{a}{2})$ at $t_{1} = \frac{a}{2}$
and that of $\zeta(2t_{1}-a)$ at $t_{1} = \frac{a}{2} + \frac{1}{2}$.
One might think that $\Gamma(t_{1}-\frac{a}{2}) \zeta(2t_{1}-a)$
could have an infinite number of poles at $t_{1} = \frac{a}{2} - n\ (n=1,2,3\cdots)$.
This is not, however, the case because of the property 
$\zeta(-2n) = 0$ for $n=1,2,3,\cdots$.
This is an important observation, which will be used throughout our analyses.

The $t_{1}$-integration on the left-hand-side of Eq.(\ref{eqA.5}) can be performed
as the residue integral and the result is
%
\begin{align}
&\hspace{-5mm}
 \frac{1}{2\pi i}\int_{c_{1}-i\infty}^{c_{1}+i\infty}dt_{1}\,
  \Gamma(t_{1}-t+\tfrac{1}{2}) \zeta(2t_{1}-2t+1) \Gamma(t_{1}) \zeta(2t_{1})
   \bigg(\pi\frac{L_{1}}{L_{0}}\bigg)^{-2t_{1}} 
    \notag\\
&= \zeta(0) \Gamma(t-\tfrac{1}{2}) \zeta(2t-1)
     \bigg(\pi\frac{L_{1}}{L_{0}}\bigg)^{-2(t-\frac{1}{2})}
   + \frac{1}{2}\Gamma(\tfrac{1}{2}) \Gamma(t) \zeta(2t) 
      \bigg(\pi\frac{L_{1}}{L_{0}}\bigg)^{-2t}
       \notag\\
&\hspace{5mm}
  + \zeta(0) \Gamma(\tfrac{1}{2}-t) \zeta(1-2t)
   + \half \Gamma(\tfrac{1}{2}) \Gamma(1-t) \zeta(2-2t)
      \bigg(\pi\frac{L_{1}}{L_{0}}\bigg)^{-1},
\label{eqA.6}
\end{align}
%
which leads to Eq.(\ref{eqA.5}), as can be confirmed by direct calculations.

%
%
%
\subsection{Multiple mode summations and pole structure}
\label{sec:modesumT}
%
%
In this section, we generalize the previous analysis of the double mode summations
to the multiple ones
%
\begin{align}
&S^{(n)}(t; L_{0},L_{1},\cdots,L_{n-1}) \notag\\
&\equiv \Gamma(t) \sum_{m_{0}=1}^{\infty}\sum_{m_{1}=1}^{\infty}\cdots
         \sum_{m_{n-1}=1}^{\infty}
          \big\{ (m_{0}L_{0})^{2}+(m_{1}L_{1})^{2} + \cdots
                 + (m_{n-1}L_{n-1})^{2} \big\}^{-t}.
\label{eqA.7}
\end{align}
%
By use of the formula (\ref{appshiki1}), 
$S^{(n)}(t; L_{0},L_{1},\cdots,L_{n-1})$ can be expressed into a recursive
form as
%
\begin{align}
&S^{(n)}(t; L_{0},L_{1},\cdots,L_{n-1}) \notag\\
&= -\frac{1}{2} S^{(n-1)}(t; L_{1},L_{2},\cdots,L_{n-1})
    + \frac{\sqrt{\pi}}{2L_{0}} S^{(n-1)}(t-\tfrac{1}{2}; L_{1},L_{2},\cdots,L_{n-1})
     \notag\\
&\hspace{5mm}
  + \frac{1}{\sqrt{\pi}} \frac{1}{2\pi i}
     \int_{c_{1}-i\infty}^{c_{1}+i\infty}dt_{1}\,
      \Gamma(t_{1}-t+\tfrac{1}{2}) \zeta(2t_{1}-2t+1)
      \Big(\frac{\pi}{L_{0}}\Big)^{-2t_{1}+2t}
       S^{(n-1)}(t_{1}; L_{1},L_{2},\cdots,L_{n-1})\,,
\label{eqA.8}
\end{align}
%
where we have defined
%
\begin{align}
&S^{(n-k)}(t; L_{k},L_{k+1},\cdots,L_{n-1}) \notag\\
&= \Gamma(t) \sum_{m_{k}=1}^{\infty}\sum_{m_{k+1}=1}^{\infty}\cdots
         \sum_{m_{n-1}=1}^{\infty}
          \big\{ (m_{k}L_{k})^{2}+(m_{k+1}L_{k+1})^{2} + \cdots
                 + (m_{n-1}L_{n-1})^{2} \big\}^{-t}
\label{eqA.9}
\end{align}
%
for $k=0,1,\cdots,n-1$.

In the following, we shall show that the third term in Eq.(\ref{eqA.8})
has the property
%
\begin{align}
&\frac{1}{2\pi i}\int_{c-i\infty}^{c+i\infty}dt\,
 f(t)\,\frac{1}{2\pi i}\int_{c_{1}-i\infty}^{c_{1}+i\infty}dt_{1}\,
  \Gamma(t_{1}-t+\tfrac{1}{2}) \zeta(2t_{1}-2t+1) 
   \Big(\frac{\pi}{L_{0}}\Big)^{-2t_{1}+2t}
    \notag\\
&\hspace{10mm}    
    \times S^{(n-1)}(t_{1}; L_{1},L_{2},\cdots,L_{n-1})
    = 0\,,
\label{eqA.10}
\end{align}
%
where $f(t)$ is any function that has no poles inside the region of the
residue integral with respect to $t$.
Eq.(\ref{eqA.10}) implies that the poles of the third term in Eq.(\ref{eqA.8})
with respect to $t$ do not totally contribute to the residue integral of t,
although the third term in Eq.(\ref{eqA.8}) has several poles with respect to $t$,
as we will see below.

By repeatedly using the relation (\ref{eqA.8}), $S^{(n)}(t; L_{0},L_{1},\cdots,L_{n-1})$
can be expressed into the form
%
\begin{align}
S^{(n)}(t; L_{0},L_{1},\cdots,L_{n-1})
 &= (P_{1}+P_{2}+P_{3})^{n-1} S^{(n)}(t; L_{0},L_{1},\cdots,L_{n-1})
    \notag\\
 &= \sum_{j_{1}=1}^{3}\sum_{j_{2}=1}^{3}\cdots \sum_{j_{n-1}=1}^{3}
    P_{j_{1}} P_{j_{2}} \cdots P_{j_{n-1}}
    S^{(n)}(t; L_{0},L_{1},\cdots,L_{n-1})\,,
\label{eqA.11}
\end{align}
%
where the operation of $P_{j}\ (j=1,2,3)$ is defined by
%
\begin{align}
P_{1}S^{(n-k)}(t; L_{k},L_{k+1},\cdots,L_{n-1})
 &= -\frac{1}{2} S^{(n-k -1)}(t; L_{k+1},L_{k+2},\cdots,L_{n-1})\,,
    \label{eqA.12}\\
P_{2}S^{(n-k)}(t; L_{k},L_{k+1},\cdots,L_{n-1})
 &= \frac{\sqrt{\pi}}{2L_{k}} S^{(n-k -1)}(t-\tfrac{1}{2}; L_{k+1},L_{k+2},\cdots,L_{n-1})\,,
    \label{eqA.13}\\
P_{3}S^{(n-k)}(t; L_{k},L_{k+1},\cdots,L_{n-1})
 &= \frac{1}{\sqrt{\pi}}\frac{1}{2\pi i}\int_{c_{1}-\infty}^{c_{1}+i\infty}dt_{1}\,
     \Gamma(t_{1}-t+\tfrac{1}{2}) \zeta(2t_{1}-2t+1)
      \Big(\frac{\pi}{L_{k}}\Big)^{-2t_{1}+2t}
      \notag\\
 &\hspace{5mm} \times S^{(n-k -1)}(t_{1}; L_{k+1},L_{k+2},\cdots,L_{n-1})\,.
\label{eqA.14}
\end{align}
%

For instance, let us consider the term 
$(P_{1})^{n-\ell-m-1} (P_{2})^{\ell} (P_{3})^{m} S^{(n)}(t; L_{0},L_{1},\cdots,L_{n-1})$,
which is explicitly given by
%
\begin{align}
&(P_{1})^{n-\ell-m-1} (P_{2})^{\ell} (P_{3})^{m} S^{(n)}(t; L_{0},L_{1},\cdots,L_{n-1})
  \notag\\
&\hspace{5mm}
 = \frac{1}{\sqrt{\pi}}\frac{1}{2\pi i}\int_{c_{1}-i\infty}^{c_{1}+i\infty}dt_{1}\,
     \Gamma(t_{1}-t+\tfrac{1}{2}) \zeta(2t_{1}-2t+1)
      \Big(\frac{\pi}{L_{0}}\Big)^{-2t_{1}+2t}
      \notag\\
&\hspace{10mm}\times
   \frac{1}{\sqrt{\pi}}\frac{1}{2\pi i}\int_{c_{2}-i\infty}^{c_{2}+i\infty}dt_{2}\,
     \Gamma(t_{2}-t_1+\tfrac{1}{2}) \zeta(2t_{2}-2t_{1}+1)
      \Big(\frac{\pi}{L_{1}}\Big)^{-2t_{2}+2t_{1}}
       \times \cdots 
   \notag\\
&\hspace{10mm}
\times
    \frac{1}{\sqrt{\pi}}\frac{1}{2\pi i}\int_{c_{m}-i\infty}^{c_{m}+i\infty}dt_{m}\,
     \Gamma(t_{m}-t_{m-1}+\tfrac{1}{2}) \zeta(2t_{m}-2t_{m-1}+1)
      \Big(\frac{\pi}{L_{m-1}}\Big)^{-2t_{m}+2t_{m-1}}
      \notag\\
&\hspace{10mm}
    \times
    \Big(\frac{\sqrt{\pi}}{2L_{m}}\Big)\Big(\frac{\sqrt{\pi}}{2L_{m+1}}\Big) \cdots
    \Big(\frac{\sqrt{\pi}}{2L_{\ell+m-1}}\Big) \Big(-\frac{1}{2}\Big)^{n-\ell-m-1}
    S^{(1)}(t_{m}-\tfrac{\ell}{2};L_{n-1})
    \notag\\
&\hspace{5mm}
 = \frac{(-1)^{n-\ell-m-1}(\sqrt{\pi})^{\ell-m}}{2^{n-m-1}}
   \frac{(L_{n-1})^{\ell}}{L_{m}L_{m+1}\cdots L_{\ell+m-1}}
   \Big(\frac{\pi}{L_{0}}\Big)^{2t}
   \notag\\
&\hspace{10mm}
   \times
   \frac{1}{2{\pi}i}\int_{c_{1}-i\infty}^{c_{1}+i\infty}dt_{1}\,
     \Gamma(t_{1}-t+\tfrac{1}{2}) \zeta(2t_{1}-2t+1)
      \Big(\frac{L_{0}}{L_{1}}\Big)^{2t_{1}}
      \notag\\
&\hspace{10mm}
    \times      
    \frac{1}{2{\pi}i}\int_{c_{2}-i\infty}^{c_{2}+i\infty}dt_{2}\,
     \Gamma(t_{2}-t_{1}+\tfrac{1}{2}) \zeta(2t_{2}-2t_{1}+1)
      \Big(\frac{L_{1}}{L_{2}}\Big)^{2t_{2}}
         \times \cdots 
   \notag\\
&\hspace{10mm}
   \times
    \frac{1}{2{\pi}i}\int_{c_{m}-i\infty}^{c_{m}+i\infty}dt_{m}\,
     \Gamma(t_{m}-t_{m-1}+\tfrac{1}{2}) \zeta(2t_{m}-2t_{m-1}+1)
     \notag\\
&\hspace{10mm}
   \times
      \Gamma(t_{m}-\tfrac{\ell}{2}) \zeta(2t_{m}-\ell)
       \Big(\frac{L_{m-1}}{\pi L_{n-1}}\Big)^{2t_{m}} .
\label{eqA.15}
\end{align}
%
In the second equality of Eq.(\ref{eqA.15}), we have used
%
\begin{align}
S^{(1)}(t_{m}-\tfrac{\ell}{2};L_{n-1})
 = \Gamma(t_{m}-\tfrac{\ell}{2})
    \sum_{m_{n-1}=1}^{\infty}(m_{n-1}L_{n-1})^{-2t_{m}+\ell}
 = \frac{\Gamma(t_{m}-\frac{\ell}{2}) \zeta(2t_{m}-\ell)}{(L_{n-1})^{2t_{m}-\ell}}.
\label{eqA.16}
\end{align}
%

Let $f(t)$ be any function which has no pole inside the region of the residue
integral with respect to $t$.
Then, we can show that
%
\begin{align}
\frac{1}{2{\pi}i}\int_{c-i\infty}^{c+i\infty}dt\,f(t)
 \big[\,
      (P_{1})^{n-\ell-m-1} (P_{2})^{\ell} (P_{3})^{m} S^{(n)}(t; L_{0},L_{1},\cdots,L_{n-1})
      \,\big]
   = 0
\label{eqA.17}
\end{align}
%
if $m \ge 1$.

To show Eq.(\ref{eqA.17}), we perform the residue integrals of Eq.(\ref{eqA.15})
with respect to $\{ t_{m}, t_{m-1}$, $\cdots, t_{2} \}$ successively by use of the
relations
%
\begin{align}
&\frac{1}{2{\pi}i}\int_{c_{j}-i\infty}^{c_{j}+i\infty}dt_{j}\,g(t_{j})
  \Gamma(t_{j}-t_{j-1}+\tfrac{1}{2}) \zeta(2t_{j}-2t_{j-1}+1)
   \Gamma(t_{j}- \tfrac{k}{2}) \zeta(2t_{j}-k)
    \notag\\
&
 = g(t_{j-1}-\tfrac{1}{2}) \zeta(0) 
    \Gamma(t_{j-1}-\tfrac{k+1}{2}) \zeta(2t_{j-1}-(k+1))
     + \frac{1}{2} g(t_{j-1}) \Gamma(\tfrac{1}{2})
         \Gamma(t_{j-1}-\tfrac{k}{2}) \zeta(2t_{j-1}-k)
         \notag\\
&\hspace{5mm}
   + g(\tfrac{k}{2}) \zeta(0) 
      \Gamma(\tfrac{k+1}{2}-t_{j-1}) \zeta(k+1-2t_{j-1})
     + \frac{1}{2} g(\tfrac{k+1}{2}) \Gamma(\tfrac{1}{2})
         \Gamma(\tfrac{k+2}{2}-t_{j-1}) \zeta(k+2-2t_{j-1})\,,
   \label{eqA.18}\\
&\frac{1}{2{\pi}i}\int_{c_{j}-i\infty}^{c_{j}+i\infty}dt_j\,g(t_{j})
  \Gamma(t_{j}-t_{j-1}+\tfrac{1}{2}) \zeta(2t_{j}-2t_{j-1}+1)
   \Gamma(\tfrac{k}{2}-t_{j}) \zeta(k-2t_{j})
    \notag\\
&
 = g(t_{j-1}-\tfrac{1}{2}) \zeta(0) 
    \Gamma(\tfrac{k+1}{2}-t_{j-1}) \zeta(k+1-2t_{j-1})
     + \frac{1}{2} g(t_{j-1}) \Gamma(\tfrac{1}{2})
         \Gamma(\tfrac{k}{2}-t_{j-1}) \zeta(k-2t_{j-1})
         \notag\\
&\hspace{5mm}
   - g(\tfrac{k}{2}) \zeta(0) 
      \Gamma(\tfrac{k+1}{2}-t_{j-1}) \zeta(k+1-2t_{j-1})
     - \frac{1}{2} g(\tfrac{k-1}{2}) \Gamma(\tfrac{1}{2})
         \Gamma(\tfrac{k}{2}-t_{j-1}) \zeta(k-2t_{j-1})\,,
\label{eqA.19}
\end{align}
%
where $g(t_{j})$ is any function that has no poles inside the region of the
residue integral with respect to $t_{j}$.
We note that there appears only the combination of the type 
$\Gamma(\pm(t_{j-1}-\frac{\ell}{2})) \zeta(\pm(2t_{j-1}-\ell))$
for some integer $\ell$ on the right-hand-side of the formulas (\ref{eqA.18}) 
and (\ref{eqA.19}), so that we can perform the residue integrals with respect to
$t_{m}, t_{m-1}, \cdots, t_2$, successively.

After performing the residue integrals with respect to $t_{j}\ (j=m,m-1,\cdots,2)$
by use of the formulas (\ref{eqA.18}) and (\ref{eqA.19}),
Eq.(\ref{eqA.15}) can be written as the sum of the terms proportional to the
following types of the integrals:
%
\begin{align}
\frac{1}{2{\pi}i}\int_{c_{1}-i\infty}^{c_{1}+i\infty}dt_{1}\,g_{\ell}(t_{1})
 \Gamma(t_{1}-t+\tfrac{1}{2}) \zeta(2t_{1}-2t+1)
  \Gamma\big(\pm(t_{1}-\tfrac{\ell}{2})\big) \zeta\big(\pm(2t_{1}-\ell)\big),
\label{eqA.20}
\end{align}
%
where $\ell$ is some positive integer and $g_{\ell}(t_{1})$ is some function,
which has no poles inside the region of the residue integral with respect $t_{1}$.

It is now easy to show Eq.(\ref{eqA.17}), which follows from the relation
%
\begin{align}
&\frac{1}{2{\pi}i}\int_{c-i\infty}^{c+i\infty}dt\,f(t)
 \frac{1}{2{\pi}i}\int_{c_{1}-i\infty}^{c_{1}+i\infty}dt_{1}\,g_{\ell}(t_{1})
  \Gamma(t_{1}-t+\tfrac{1}{2}) \zeta(2t_{1}-2t+1)
   \notag\\
&\hspace{10mm}  
  \times \Gamma\big(\pm(t_{1}-\tfrac{\ell}{2})\big) \zeta\big(\pm(2t_{1}-\ell)\big)
   = 0
\label{eqA.21}
\end{align}
%
with Eqs.(\ref{eqA.18}) and (\ref{eqA.19}).
It should be emphasized that Eq.(\ref{eqA.21}) does not mean that the
integrand of Eq.(\ref{eqA.17}) has no poles with respect to $t$.
In fact, the integrand of Eq.(\ref{eqA.17}) has several poles with respect to
$t$ and each pole contributes to the residue integral of $t$, though the
sum of their residues totally cancels each other.

We have proved Eq.(\ref{eqA.17}) for a special order of 
$P_{j_{1}}P_{j_{2}}\cdots P_{j_{n-1}}$.
Since $P_{j}P_{k} S^{(n-k)}(t;L_{k},L_{k+1},$
$\cdots,L_{n-1})$ $(j,k=1,2,3)$
is identical to the opposite order of 
$P_{k}P_{j} S^{(n-k)}(t;L_{k},L_{k+1},\cdots,L_{n-1})$
with the exchange of $L_{k} \leftrightarrow L_{k+1}$
(and with a shift of the integration parameter, if necessary),
we generally have
%
\begin{align}
\frac{1}{2{\pi}i}\int_{c-i\infty}^{c+i\infty}dt\,f(t)
 \big[\,P_{j_{1}}P_{j_{2}}\cdots P_{j_{n-1}} S^{(n)}(t; L_{0},L_{1},\cdots,L_{n-1})
      \,\big]
   = 0
\label{eqA.22}
\end{align}
%
if some of $j_{s}\ (s=1,2,\cdots,n-1)$ take the value of $3$.
This result immediately leads to Eq.(\ref{eqA.10}) because Eq.(\ref{eqA.10})
corresponds to the case of $j_{1}=3$ in Eq.(\ref{eqA.22}).
%
%
%
%
%

\end{document}